\documentclass[useAMS,usenatbib]{mn2e}

\usepackage{amsmath}
\usepackage{amssymb}
\usepackage[dvipsnames]{xcolor}
\usepackage{graphicx}
\usepackage{graphics}
\usepackage{hyperref}
\usepackage{multirow}
\usepackage{multicol}
\usepackage{array}
\usepackage{breqn}
\newcolumntype{H}{>{\setbox0=\hbox\bgroup}c<{\egroup}@{}}
\usepackage{comment}
\usepackage{ulem}
\usepackage[T1]{fontenc}

\hypersetup{
     colorlinks   = true,
     citecolor    = blue
}

\def\apj{{ ApJ}}
\def\apjl{{ApJL}}
\def\apjs{{ ApJS}}

\def\aap{{ A\&A}}
\def\aj{{ AJ}}
\def\mnras{{ MNRAS}}
\def\araa{{ ARA\&A}}

\def\actaa {{ Acta Astronomica}}

\def\pasj{{ PASJ}}
\def\pasa{{PASA}}

\def\sovast{{ Sov. Astron.}}
\def\apspr{{ Astrophysics and Space Physics Reviews}}

\def\mr{\mathrm}
\def\mb{\mathbf}
\def\mc{\mathcal}
\def\hx{\hat{\mb{x}}}

\def\hz{\hat{\mb{z}}}
\def\d{\mr{d}}

\def\Msun{M_\odot}
\def\Rsun{R_\odot}
\def\Rd{R_{\rm d}}
\def\Rsph{R_{\rm sph}}
\def\Rc{R_{\rm c}}
\def\Md{M_{\rm d}}

\def\Rs{R_{\rm s}}
\def\kaps{\kappa_{\rm s}}
\def\kapa{\kappa_{\rm a}}
\def\vw{v_{\rm w}}
\def\fL2{f_{\rm L2}}
\def\kB{k_{\rm B}}

\def\OmgK{\Omega_{\rm K}}

\def\mp{m_{\rm p}}
\def\cs{c_{\rm s}}

\def\veq{v_{\rm eq}}

\def\rhoeq{\rho_{\rm eq}}

\def\theobs{\Theta_{\rm obs}}
\def\Rg{R_{\rm g}}

\def\msunyr{{M_\odot\,\mr{yr}^{-1}}}

\def\fOmg{{f_{\Omega}}}

\newcommand{\note}[1]{{#1}}
\newcommand{\lrb}[1]{\left({#1}\right)}
\newcommand{\lrsb}[1]{\left[{#1}\right]}

\newcommand{\myemail}{wenbinlu@berkeley.edu}

\title[L2 mass loss]{On rapid binary mass transfer --- I. Physical model}
\author[Lu, Fuller, Quataert \& Bonnerot]
  {Wenbin Lu$^{1, 2}$\thanks{\myemail}, Jim Fuller$^3$, Eliot Quataert$^2$, Cl{\'e}ment Bonnerot$^{4}$\\
  $^1$Departments of Astronomy and Theoretical Astrophysics Center, UC Berkeley, Berkeley, CA 94720, USA\\
  $^2$Department of Astrophysical Sciences, Princeton University, Princeton, NJ 08544, USA\\
  $^3$TAPIR, Walter Burke Institute for Theoretical Physics, Mail Code
  350-17, Caltech, Pasadena, CA 91125, USA\\
  $^4$Niels Bohr International Academy, Niels Bohr Institute, Blegdamsvej 17, DK-2100 Copenhagen Ø, Denmark
  }

\begin{document}
\label{firstpage}
\maketitle

\begin{abstract}
In some semi-detached binary systems, the donor star may transfer mass to the companion at a very high rate. We propose that, at sufficiently high mass-transfer rates such that the accretion disk around the companion becomes geometrically thick (or advection-dominated) near the disk outer radius, a large fraction of the transferred mass may be lost through the outer Lagrangian (L2) point, as a result of the excessive energy generated by viscous heating that cannot be efficiently radiated away. A physical model is constructed where the L2 mass loss fraction is given by the requirement that the remaining material in the disk has Bernoulli number equal to the L2 potential energy. Our model predicts significant L2 mass loss at mass transfer rates exceeding $\mbox{a few}\, 10^{-4}\,\msunyr$. An equatorial circum-binary outflow (CBO) is formed in these systems. Implications for the orbital evolution and the observational appearance of the system are discussed. In particular, (1) rapid angular momentum loss from the system tends to shrink the orbit and hence may increase the formation rate of mergers and gravitational-wave sources; and (2) photons from the hot disk wind are reprocessed by the CBO into longer wavelength emission in the infrared bands, consistent with \textit{Spitzer} observations of some ultra-luminous X-ray sources.
\end{abstract}

\section{introduction}

Binary mass transfer is one of the key aspects of stellar evolution and has been extensively studied in the past few decades \citep{paczynski71_binary_evolution}. Observational constraints on the physics involved in the process come from the large number of systems that are undergoing or about to undergo mass transfer, as well as many other objects that experienced mass transfer in the past. Perhaps the least well-understood cases are the short-lived systems where the donor transfers mass to the companion at rates higher than about $10^{-4}\,\msunyr$ and large uncertainties exist in predicting the fate of these systems \citep{podsiadlowski92_pre-SN_evolution, langer12_massive_star_evolution, ivanova13_common_envelope, postnov14_binary_evolution}.

Some physical examples potentially reaching such high mass transfer rates are: (1) the Galactic micro-quasar SS433 \citep{fabrika04_SS433} and some extragalactic ultra-luminous X-ray sources \citep[ULXes,][]{kaaret17_ULXs}; (2) binaries where the donor transfers mass to the companion on the Kelvin-Helmholtz timescale of its envelope, such as massive $(\gtrsim 10\Msun)$ Hertzsprung-gap stars undergoing envelope expansion in a binary with period less than $10^3\rm\, d$ \citep{vandenheuvel17_BBH_stable_mass_transfer, marchant21_L2massloss, klencki21_mass_transfer_rate}, and short-period (less than $2\,$day) helium star-neutron star (NS) binaries prior to the formation of merging double NSs \citep{tauris15_ultra_stripped_SN};
(3) binaries undergoing unstable  overflow just before the common envelope (CE) phase or stellar merger \citep{paczynski72_unstable_mass_transfer, hjellming87_unstable_mass_transfer, soberman97_unstable_mass_transfer, ge10_MT_stability, pavlovskii17_direct_L2L3_MT, metzger21_CV_mergers}, and (4) pre-supernova binaries where one of the stars may rapidly expand in the final stages of nuclear burning weeks to decades before the explosion \citep{quataert12_wave_driven_outburst, mcley14_wave_driven_envelope_expansion, fuller17_wave_driven_outburst, wu21_wave_driven_outbursts}. Gravitational wave sources detected in the past few years provide new constraints on the mass transfer physics by sampling the end products of some of these systems \citep[e.g.,][]{abbott20_bNS_rate, abbott21_BBH_statistics}.

The goal of this paper to study the hydrodynamics of the transferred mass by modeling the super-Eddington accretion disk around the companion as well as the interaction between the material driven away from the disk and the binary Roche potential. We propose that, at mass-transfer rates exceeding a critical value (to be calculated in this work) such that the accretion disk around the companion becomes geometrically thick (or advection-dominated) near the disk outer radius, an order unity fraction of the transferred mass may be lost from the system through the outer Lagrangian (L2) point. This is because the advection-dominated accretion flow in the outer disk is energetically capable of driving material to L2 equipotential surface and beyond \citep{narayan94_ADAF, narayan95_ADAF}. We construct a physical model for the accretion disk and calculate the fraction of the transferred mass that is lost through the L2 point.

Binary mass transfer at such a high rate has been previously considered by \citet{king99_disk_outflow} and \citet{begelman06_superEddington_wind}, who propose that the majority of the transferred mass is blown away to infinity in the form of a fast, super-Eddington wind. However, because lifting material to the L2 potential is much less energetically demanding than to infinity, it is likely that the system takes the more energetically efficient solution, which is described in this paper. Evidence for such a solution has been seen in numerical simulations by \citet{bisikalo98_equatorial_mass_loss, sytov07_L2_loss_common_envelope, macleod18_L2_loss_common_envelope, macleod18_L2_loss_binary_merger} for systems right before the onset of the CE phase, although these works did not treat the viscous accretion onto the companion. Another prescription used in binary population synthesis calculations is that when the mass-transfer exceeds a critical rate, the system undergoes CE evolution \citep{ivanova03_CE_at_high_Mdot}. Our model allows explicit calculation of the binary orbital evolution taking into account the angular momentum carried away by L2 mass loss. This does not necessarily lead to CE. Only the cases with extremely rapid orbital shrinkage on a dynamical time may undergo CE evolution.



Observationally, the binary system SS433 in our Galaxy, with a mass-transfer rate of the order $10^{-4}\,\msunyr$ \citep{fabrika04_SS433}, indeed appears to be undergoing L2 mass loss. Studies of the optical emission line profiles and spatially resolved radio/infrared images show the existence of equatorial, circumbinary outflowing material \citep[e.g.,][]{filippenko88_ss433_disk_emission, paragi99_SS433_CBO_radio_image, blundell01_ss433_eq_outflow, blundell08_ss433_eq_outflow, waisberg19_ss433_eq_outflow}. Many nearby ($\lesssim 10\rm\, Mpc$) ULXes have optical and infrared (IR) counterparts \citep{tao11_opt_counterparts, gladstone13_opt_counterpart, heida14_NIR_counterparts}. The spectral energy distributions (SEDs) of some of these sources show IR excess far above the power-law extrapolation from the optical bands \citep{heida14_NIR_counterparts, lopez17_NIR_counterparts, lau17_MIR_counterparts, lau19_ULX_MIR}, and these authors have suggested that the IR excess might be due to a red supergiant star (or supergiant Be star) donor or circum-stellar dust. In our model, a system undergoing L2 mass loss forms an equatorially concentrated circum-binary outflow (CBO), and we show that the observed IR excess is consistent with reprocessing of the disk wind emission by the CBO.

This paper is organized as follows. We present the model for the accretion disk at different mass-transfer rates and binary separations, and then determine the fate of the transferred mass in \S \ref{sec:disk_model}. Then in \S \ref{sec:orbital_evolution}, the effect of L2 mass loss on the binary orbital evolution is discussed. In \S \ref{sec:EM_signatures}, we calculate the radiative appearance of a system undergoing L2 mass loss, focusing on the reprocessing of the disk wind emission by the circum-binary material. We discuss the limitations of our model in \S \ref{sec:discussion}. A summary of our results is provided in \S \ref{sec:summary}. The logarithm of base 10 is denoted as $\log$ throughout the paper.
\section{Fate of the Overflowing Mass}\label{sec:disk_model}
In this section, we first review the standard Roche-lobe geometry and define the parameters of the problem in \S \ref{sec:roche-lobe}. Then, we present the model for the accretion disk around the accretor including possible L2 mass loss. Two different regimes of mass-transfer rates are discussed in \S \ref{sec:outer-disk} (high $|\dot{M}_1|$) and \S \ref{sec:inner-disk} (low $|\dot{M}_1|$).

\subsection{Standard Roche-lobe Geometry}\label{sec:roche-lobe}
Let us consider a primary donor star of mass $M_1$ with a given density profile and a point-mass companion of mass $M_2$ in a circular orbit. The separation between the centers of the two stars is $a$. We define the mass ratio as
\begin{equation}
  \label{eq:9}
  q = {M_2\over M_1}, \mbox{ and set } \mu = {q\over 1 + q}. 
\end{equation}
Under the assumption that both stars are in synchronous rotation at the Keplerian angular frequency and that nearly all the mass of the primary is concentrated near its center, the Roche potential $\Phi(\mb{r})$ is given by
\begin{equation}
  \label{eq:8}
  {-a\,\Phi(\mb{r}) \over G(M_1+M_2)} ={1-\mu\over |\mb{r}|} + {\mu\over |\mb{r}-\hx|} +
  {\left(x -\mu\right)^2 + y^2\over 2},
\end{equation}
where we have placed the origin of the Cartesian coordinate system at the center of the primary star, with $\hx$ pointing towards the secondary and $\hz$ towards the direction of the angular momentum vector, and the coordinates have been normalized by the orbital separation $a$. The position of the secondary is $\mb{r}_2 = \hx$ and the center of mass of the binary system is located at $\mb{r}_{\rm CM} = \mu\hx$. The positions of stationary points are the simultaneous solutions of $\partial_x\Phi = \partial_y\Phi = \partial_z\Phi = 0$. The three Lagrangian points along the $\hx$ axis are obtained by solving the two equations above while setting $y=z=0$, and this gives
\begin{equation}
  \label{eq:10}
  (1-\mu) {x\over |x|^3} + \mu {x-1\over |x-1|^3} - x + \mu = 0.
\end{equation}
The three solutions are denoted as $x_{\rm L1}$, $x_{\rm L2}$, $x_{\rm L3}$, and the corresponding dimensionless potentials are denoted as $\Phi_{\rm L1}$, $\Phi_{\rm L2}$, $\Phi_{\rm L3}$. \note{We find the following quadratic polynomial fits to the numerical solutions}
\begin{equation}\label{eq:L1L2positions}
\begin{split}
    x_{\rm L1} &\approx -0.0355(\mr{log}\,q)^2 -0.251 \,\mr{log}\,q + 0.500,\\
    x_{\rm L2} &\approx 0.0756(\mr{log}\,q)^2 +0.424 \,\mr{log}\,q + 1.699,
\end{split}
\end{equation}
which apply for $0.01<q<1$ with fractional errors $\lesssim 0.3\%$ and \note{are slightly more accurate than linear fits} \citep[e.g.,][]{plavec64_Roche_tables,frank02_accretion_disk_book}. By symmetry, the Lagrangian positions for $q>1$ are $1-x_{\rm L1}(1/q)$ and $1-x_{\rm L2}(1/q)$.

The radius $R_{\rm v}$ of a volume-equivalent sphere for each equipotential surface (denoted by its dimensionless potential $\Phi_{\rm p}$) around the primary star can be obtained in the following way. The potential gradient on the equipotential surface far from the L1 nozzle is dominated by the gravity of the primary, $\d \Phi_{\rm p}/\d R_{\rm v} \approx (1+q)^{-1}R_{\rm v}^{-2}$, and this can be combined with the boundary condition $R_{\rm v}(\Phi_{\rm L1})=R_{\rm v, L1}$ to obtain
\begin{equation}
  \label{eq:40}
  R_{\rm v}(\Phi_{\rm p}) \approx {1\over R_{\rm
v,L1}^{-1} - (1+q)(\Phi_{\rm p}-\Phi_{\rm L1})},
\end{equation}
where the volume-equivalent radius of the Roche lobe is given by \citep{eggleton83_RvL1}
\begin{equation}
  \label{eq:16}
  R_{\rm v, L1} \approx {0.49 \over 0.6 + q^{2/3}\,\mr{ln}(1 + q^{-1/3})}.
\end{equation}

\begin{figure}
  \centering
\includegraphics[width = 0.49\textwidth]{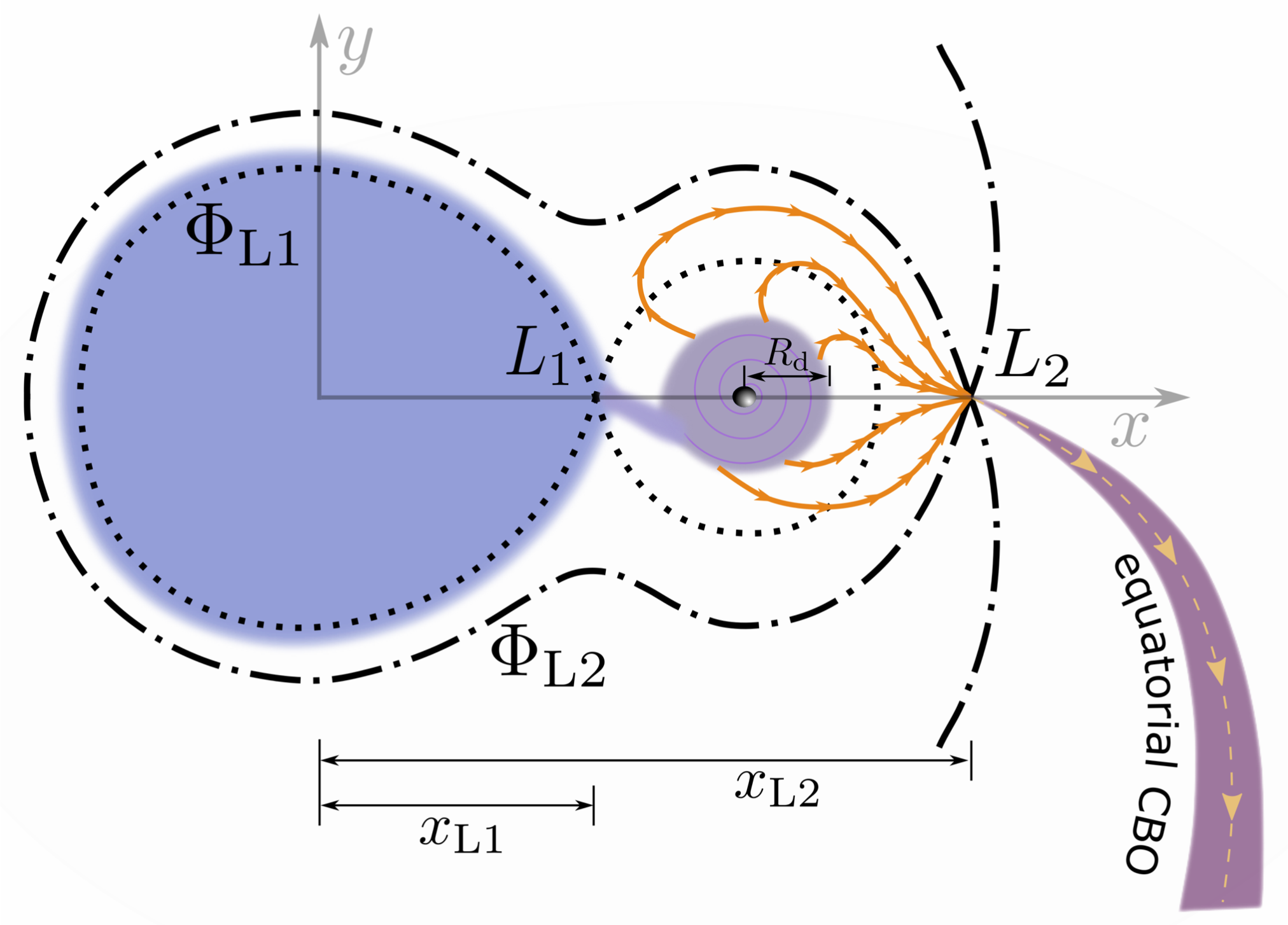}
\caption{A broad-brush picture of the model. Two equipotential surfaces with potential energies $\Phi_{\rm L1}$ (dotted curve) and $\Phi_{\rm L2}$ (dashed-dotted curve) are shown in the corotating frame of the binary system. The primary star (left) fills up its Roche lobe and is transferring mass towards the secondary (right) through the L1 nozzle. When the mass-transfer rate exceeds $\mbox{a few}\,10^{-4}\,M_\odot\rm\,yr^{-1}$, a significant fraction of the transferred mass is lost through the L2 nozzle likely in the form of a supersonic stream. The stream from the L2 nozzle expands in width as it propagates to large distances from the binary system and forms an axisymmetric equatorially concentrated circum-binary outflow (CBO). The thick orange solid lines with arrows pointing from the accretion disk to the L2 point are only for schematic purpose, whereas our crude model does not provide an accurate description of the gas in this region.
}\label{fig:sketch}
\end{figure}

A schematic picture of the model is shown in Fig. \ref{fig:sketch} for the case of $q < 1$ (the donor being more massive). When the primary star overfills its Roche lobe, mass starts to overflow towards the secondary at a rate $|\dot{M_1}|$ in the form of a super-sonic stream.
We only consider the cases where the mass ratio is not extreme ($q$ not much less or much greater than one), so $R_{\rm v}(\Phi_{\rm L2})$ is larger than $R_{\rm v}(\Phi_{\rm L1})$ by 10s of percent. This means that, if the primary manages to expand beyond the $\Phi_{\rm L2}$ surface, the Roche lobe (or the $\Phi_{\rm L1}$ surface) is deeply inside the star where the density and pressure are very high, so the mass-transfer rate towards the secondary is much larger than that directly escaping from the L2 point \citep{pavlovskii17_direct_L2L3_MT, marchant21_L2massloss}. Thus, in the following, we focus on the mass flow towards the secondary through the L1 nozzle (in between equipotential surfaces of $\Phi_{\rm L1}$ and $\Phi_{\rm L2}$), and our goal is to study the condition under which a significant fraction of the transferred mass through the L1 nozzle is lost beyond the L2 point.

In the case of $q > 1$ (the accretor being more massive\footnote{Note that, in the case of $q > 1$, we still denote the donor as the primary star with mass $M_1$ and the accretor as the secondary star with mass $M_2$.}), the L2 point is on the donor side and the L3 point is on the accretor side. In this work, we assume that the material can still be driven away from the accretion disk to reach the L2 point, but it must return to the donor side and fill the space in between the L1 and L2 equipotential surface. A potential concern is the counter-streaming motion above the donor's surface where the flow is subjected to the Kelvin-Helmhotz instability. Alternatively, the disk material with excessive energy (above $\Phi_{\rm L3}$) may leave the system from the nearby L3 nozzle. This alternative scenario of L3 mass loss likely occurs for the nearly equal-mass case with $q\approx 1$ and $q>1$ \citep[see Fig. 2 of][]{sytov07_L2_loss_common_envelope}. Nevertheless, even in the case of L3 mass loss, the model presented in this paper should give qualitatively similar results, since the difference $\Phi_{\rm L3}-\Phi_{\rm L2}$ is much smaller than the potential energy at the accretion disk radius $|\Phi(\Rd)|$ (defined in eq. \ref{eq:46}). For instance, when $q=2$, we have $\Phi_{\rm L3}-\Phi_{\rm L2} \approx 0.11 \Phi_0$ whereas $|\Phi(\Rd)| \approx 2.48\Phi_0$, where $\Phi_0=G(M_1 + M_2)/a$.

\note{We do not consider the regime where the accretor is much more massive than the donor ($q\gg 1$) because, in this limit, the L2 equipotential surface only has a very small lobe on the donor side. This means that the accretion disk and its outflow are nearly unaffected by the donor's gravity. It is possible that the majority of the mass driven away from the outer disk forms a quasi-spherical structure \citep[similar to that found by][]{bobrick17_WDNS_binary} instead of an equatorially concentrated CBO. Our model should be applicable for mass ratios from $q\ll 1$ up to $q\sim\,3$. }



\subsection{L2 Mass Loss from the Outer Accretion Disk}\label{sec:outer-disk}

The super-sonic stream leaving the L1 nozzle has specific angular momentum with respect to the secondary $\ell \approx (1-x_{\rm L1})^2 a^2 \Omega$ (where $\Omega = \sqrt{G(M_1+M_2)/a^3}$). This is an approximation obtained under the assumption that the inertial-frame acceleration of the fluid elements in the stream is dominated by the secondary's gravity. This angular momentum corresponds to a Keplerian circularization radius \citep{warner95_CV_book}
\begin{equation}
  \label{eq:41}
  R_{\rm  c}= \ell^2/GM_2 \approx (1-x_{\rm L1})^4a/\mu.
\end{equation}
If the radius of the secondary is greater than $R_{\rm c}$, then the transferred mass mostly lands on the accretor's surface. However, this only happens in the situation where the secondary is also close to filling its own Roche lobe (e.g., for mass transfer between two main-sequence stars or two white dwarfs of comparable masses).

In this paper, we restrict our analysis to the case where the secondary is smaller than the circularization radius, $R_2< R_{\rm c}$ and an accretion disk forms. The outer edge of the disk viscously spreads to a slightly larger radius $R_{\rm d, max}$ where it is tidally truncated. The tidal truncation radius $R_{\rm d,max}$ is comparable to (about $80\%$ of) the volume-equivalent radius of the Roche lobe of the secondary \citep[see][for considerations based on orbit crossing]{paczynski77_disk_size, hirose90_CV_disk_size}, but the majority of the disk mass is at smaller radii closer to $R_{\rm c}$. In the following, we consider a simplified one-zone model for the outer accretion disk at radius
\begin{equation}
  \label{eq:42}
  \Rd\simeq \Rc.
\end{equation}
\note{The potential energy at radius $\Rd$ from the accretor can be obtained from eq. (\ref{eq:8}) and, when high-order terms $\mc{O}(\Rd^2/a^2)$ are ignored, the result can be simplified into}
\begin{equation}
  \label{eq:46}
\begin{split}
  \Phi(\Rd) &\approx -{G(M_1+M_2)\over a}\left[1-\mu + {\mu a\over \Rd} +
    {1\over 2}(1-\mu)^2\right] \\
    &= -{GM_2\over \Rd}\left[1 +
    {(1-\mu)(3-\mu)\over 2\mu} {\Rd \over a}\right].
\end{split}
\end{equation}
The goal of this subsection is to understand the thermodynamic state and the fate of the disk material near $\Rd$. In particular, we show that, when the mass-transfer rate is sufficiently high such that $GM_2|\dot{M}_1|/\Rd$ is comparable to the Eddington luminosity of the secondary, a large fraction of the transferred mass is lost through the L2 nozzle.

Because $\Rd$ is about 1/3 of the volume-equivalent radius of the secondary's Roche lobe \citep{warner95_CV_book}, the gravity on the disk gas is dominated by the secondary. The vertical pressure scale height $H$ is given by the balance between pressure gradient and gravity in the direction perpendicular to the disk mid-plane
\begin{equation}
  \label{eq:disk_height}
\begin{split}
  H &= \cs/\OmgK,\ \cs=\sqrt{P/\rho},\ \OmgK =
  \sqrt{GM_2/\Rd^3}, 
\end{split}
\end{equation}
where $\OmgK$ is the Keplerian frequency, $\rho$ is the density, $\cs$ is the isothermal sound speed, and the total pressure $P$ includes the contributions from gas $P_{\rm g} = \rho \kB T/\mu_{\rm g}\mp$ ($\kB$, $\mp$ being the Boltzmann constant and proton mass) and radiation pressure $P_{\rm rad}=a_{\rm r}T^4/3$ ($a_{\rm r}$ being the radiation density constant). The gas pressure contains the mean molecular weight $\mu_{\rm g}$ which depends on the composition and ionization of the gas: $\mu_{\rm g}$ is $1.3/2.4\simeq 0.54$ for fully ionized solar abundance gas (with He-to-H number ratio of 0.1), $4/3$ (or 2) for fully (or singly) ionized helium, and $\gtrsim 2$ for heavier composition. In this work, we take $\mu_{\rm g}=0.54$ as fiducial. The gas-radiation mixture is assumed to be in local thermodynamic equilibrium at the same temperature $T$, as a result of efficient coupling by absorption/emission as well as Compton scattering.

The dynamical evolution of the disk is driven by viscosity \citep{shakura73_alpha_disk},
\begin{equation}
  \label{eq:18}
  \nu_{\rm vis} =\alpha \cs H =  \alpha H^2 \OmgK,
\end{equation}
and the corresponding viscous time is
\begin{equation}
  \label{eq:19}
  t_{\rm vis} = \Rd^2/\nu_{\rm vis} = [\alpha (H/\Rd)^2\OmgK]^{-1},
\end{equation}
where we will take a fiducial $\alpha=0.1$.
In a quasi-steady state, the dynamical evolution of the disk is controlled by continuous mass supply from the L1 nozzle, viscous accretion towards the secondary star, and possible mass loss from the L2 nozzle. It may be tempting to include another mass loss term from a fast wind that directly escape the system \citep[e.g.,][]{piran77_disk_wind, king99_disk_outflow}. This requires acceleration of gas near $R_{\rm d}$ to above local escape speed. This is much more energetically demanding than lifting the material to the L2 potential, and we assume that the gas takes the more energetically efficiently path. 

Note that viscous accretion drives a fraction of gas to much smaller radii $R\ll R_{\rm d}$, where the Keplerian rotation velocity is much higher than the corotation speed at the L2 point, so the majority of the inner disk outflow launched from $R\ll R_{\rm d}$ is not strongly affected by the binary orbit and should escape the system as fast wind \citep{blandford99_ADIOS}. However, there could still be a small fraction of the outflow launched from the inner disk ($R\ll R_{\rm d}$) that is captured by the binary orbit and contributes to L2 mass loss. This will be discussed later in the next subsection. For the moment, we focus on the gas near radius $R_{\rm d}$.


We assume that a fraction $f_{\rm L2}$ of the mass inflow rate $|\dot{M}_1|$ is channeled through the L2 nozzle, and the accretion rate is given by
\begin{equation}
  \label{eq:20}
  (1-f_{\rm L2})|\dot{M}_1|  \simeq {\Md \over t_{\rm vis}}, \ \Md \simeq 2\pi \rho \Rd^2 H,
\end{equation}
where $\Md$ is the disk mass and the parameter $f_{\rm L2}$ will be determined later in a self-consistent way. Let us define a dimensionless scale height 
\begin{equation}
  \label{eq:29}
  \theta \equiv H/\Rd,
\end{equation}
and hence the disk density is given by
\begin{equation}
  \label{eq:47}
  \rho = {(1-\fL2)|\dot{M}_1|\over 2\pi \alpha \OmgK \Rd^3} {1\over \theta^3}.
\end{equation}
Then, equations (\ref{eq:disk_height}--\ref{eq:20}) can be combined to give a relation between $\theta$, the disk temperature $T$, and the L2 mass loss fraction $\fL2$,
\begin{equation}
  \label{eq:43}
 {c_1T^4\over 1-\fL2}\theta^3 - \theta^2 + c_2T = 0,
\end{equation}
where 
\begin{equation}\label{eq:c1c2}
    c_1 = { 2\pi a_{\rm r} \alpha
   \Rd \over 3\OmgK|\dot{M}_1|},\ c_2 = {\kB \Rd
   \over GM_2\mu_{\rm g}\mp}.
\end{equation}

On the other hand, the disk temperature is determined by energy conservation
\begin{equation}
  \label{eq:21}
  Q_{\rm vis}^+ + Q_{\rm sh}^+ = Q_{\rm rad}^{-} + Q_{\rm adv}^{-} +
  Q_{\rm L2}^{-},
\end{equation}
where the heating ($Q^+$) and cooling ($Q^{-}$) rates are defined below. We assume the disk to be rotating at the Keplerian frequency $\OmgK$. The deviation from Keplerian rotation due to radial pressure gradient is small, $\mc{O}(H^2/\Rd^2)$, and does not qualitatively affect our results.
The viscous heating rate per unit mass is
\begin{equation}
  \label{eq:22}
  {Q_{\rm vis}^+\over \Md}\simeq {9\over 4} \nu_{\rm vis}\OmgK^2 = 
 {9\over 4} {GM_2/\Rd \over t_{\rm vis}}.
\end{equation}
The incoming stream strikes the disk near the outer edge and eventually circularizes to near-Keplerian motion close to $\Rd$, and this generates a total shock heating rate of
\begin{equation}
  \label{eq:48}
  Q_{\rm sh}^+ \simeq |\dot{M}_1| \left(\Phi_{\rm L1}-\Phi(\Rd) -
  {GM_2\over 2\Rd}\right),
\end{equation}
where we have included the kinetic energy $GM_2/2\Rd$ but ignored the (subdominant) thermal energy of the circularized gas.
The radiative cooling rate is given by
\begin{equation}
  \label{eq:Qrad}
  Q_{\rm rad}^{-}\simeq 2\pi \Rd^2 {U_{\rm rad}c \over \tau/2}, \ \tau
  = \rho \kappa H,
\end{equation}
where $U_{\rm rad} = a_{\rm r}T^4$ is the radiation energy density, $c$ is the speed of light, $\tau$ is the optical depth in the vertical direction, and $\kappa$ is the Rosseland-mean opacity. We have taken the vertical diffusive radiative flux to be $U_{\rm rad}c /(\tau/2)$, where the factor of $1/2$ in the denominator is based on the consideration that most photons are generated at some height ($\sim H/2$) away from the disk mid-plane where the optical depth is reduced compared to that of the entire disk. Note that we are interested in the regime  where the disk is highly optically thick $\tau\gg 1$ and we have checked that all our solutions satisfy this condition.

As for the opacity $\kappa(\rho, T)$, we smoothly blend the high-temperature ($T\gtrsim 10^4\rm\, K$) tables from OPAL \citep{iglesias96_OPAL} and low-temperature ($10^3\lesssim T\lesssim 10^4\rm\, K$) tables by \citet[][]{ferguson05_lowT_opacity}, which include effects of molecules and dust grains. These tables are conveniently collected by the $\mathtt{MESA}$ code \citep{paxton19_mesa} in the $\texttt{/mesa/kap/}$ directory. Our fiducial gas composition is H-rich with mass fractions $X=0.7$ (H), $Y=0.28$ (He), $Z=0.02$ (metals, with solar abundance), for which the names of the opacity tables are ``gn93$\_$z0.02$\_$x0.7.data'' and ``lowT$\_$fa05$\_$gn93$\_$z0.02$\_$x0.7.data''. In the Appendix, we also show the results for H-poor gas composition with $(X=0$, $Z=0.02)$ and for a low-metallicity case $(X=0.7, Z=0.001)$. A key difference between using realistic opacity tables and analytic Kramer's opacity is that the opacity is strongly enhanced near $T\sim 2\times 10^5\rm\, K$ due to bound-bound transitions of Fe \citep{badnell05_opacity_project}, and the iron opacity bump increases the L2 mass-loss fraction.


The cooling rate per unit mass due to heat advection by the radial inflow is given by
\begin{equation}
  \label{eq:Qadv}
  {Q_{\rm adv}^{-}\over \Md} = v_{\rm r} T {\d s\over \d R} \simeq
  {3\over 2}{U\over P}
  {GM_2/\Rd\over t_{\rm vis}} \theta^2,
\end{equation}
where $s$ is the specific entropy and $v_{\rm r}\simeq -3\nu_{\rm vis}/(2\Rd)$ is the radial velocity driven by viscous angular
momentum transfer. To obtain the second expression in equation (\ref{eq:Qadv}), we have made use of $T\d s = \d h - \d P/\rho$, $\d h/\d R\simeq h/\Rd$ for the specific enthalpy $h = (U+P)/\rho$ and total energy density $U \simeq 3\rho \kB T/2\mu_{\rm g}\mp + a_{\rm r}T^4$, and $\d P/\d R\simeq P/\Rd$. Determining the precise radial gradients requires at least one-dimensional modeling for the entire disk \citep[e.g.,][]{blandford99_ADIOS, blandford04_2D_ADIOS}, and our qualitative results are only weakly affected by these numerical factors of order unity. Note that $Q_{\rm adv}^-\propto (H/\Rd)^2$ means that advective cooling is only important for a geometrically thick disk.

Finally, we argue that mass loss from the L2 nozzle occurs at sufficiently high mass-transfer rates. This is because, in the absence of L2 mass loss (setting $f_{\rm L2} = 0$ in equations \ref{eq:20} and \ref{eq:21}), if the disk is in the advection-dominated state ($Q_{\rm adv}^{-}\gg Q_{\rm rad}^{-}$), the system necessarily leads to a positive Bernoulli number, as showed by \citet{narayan94_ADAF, narayan95_ADAF} and \citet{blandford99_ADIOS}. The physical consequence is that a significant fraction of the transferred mass can be driven away from the system. The most energetically efficient way of achieving this is to push some gas slightly over the L2 nozzle, and this is subsequently accelerated by the torque from the binary and flies to very large distances \citep{shu79_L2_stream}. Recent numerical simulations of binary mass transfer with adiabatic hydrodynamics \citep{macleod18_L2_loss_binary_merger, macleod18_L2_loss_common_envelope} indeed find L2 mass loss and hence support our proposal, but future radiation magneto-hydrodynamic (MHD) simulations are needed to further test our model.

Based on the above arguments, we take the cooling rate due to L2 mass loss to be
\begin{equation}
  \label{eq:QL2}
  Q_{\rm L2}^- \simeq f_{\rm L2}|\dot{M}_1| \left(\Phi_{\rm L2} -
    \Phi(\Rd)-{GM_2\over 2\Rd}\right). 
\end{equation}
More realistically, when reaching the L2 point, the escaping gas may have specific energy higher than $\Phi_{\rm L2}$, and the residual kinetic energy and enthalpy will affect their kinematics at larger radii $R\gg a$ \citep{pejcha16_L2_stream_shape}. However, determining the residual energy of the escaping gas at the L2 point is beyond the scope of the current work as it requires detailed numerical simulations.


\begin{figure*}
  \centering
\includegraphics[width = 0.49\textwidth]{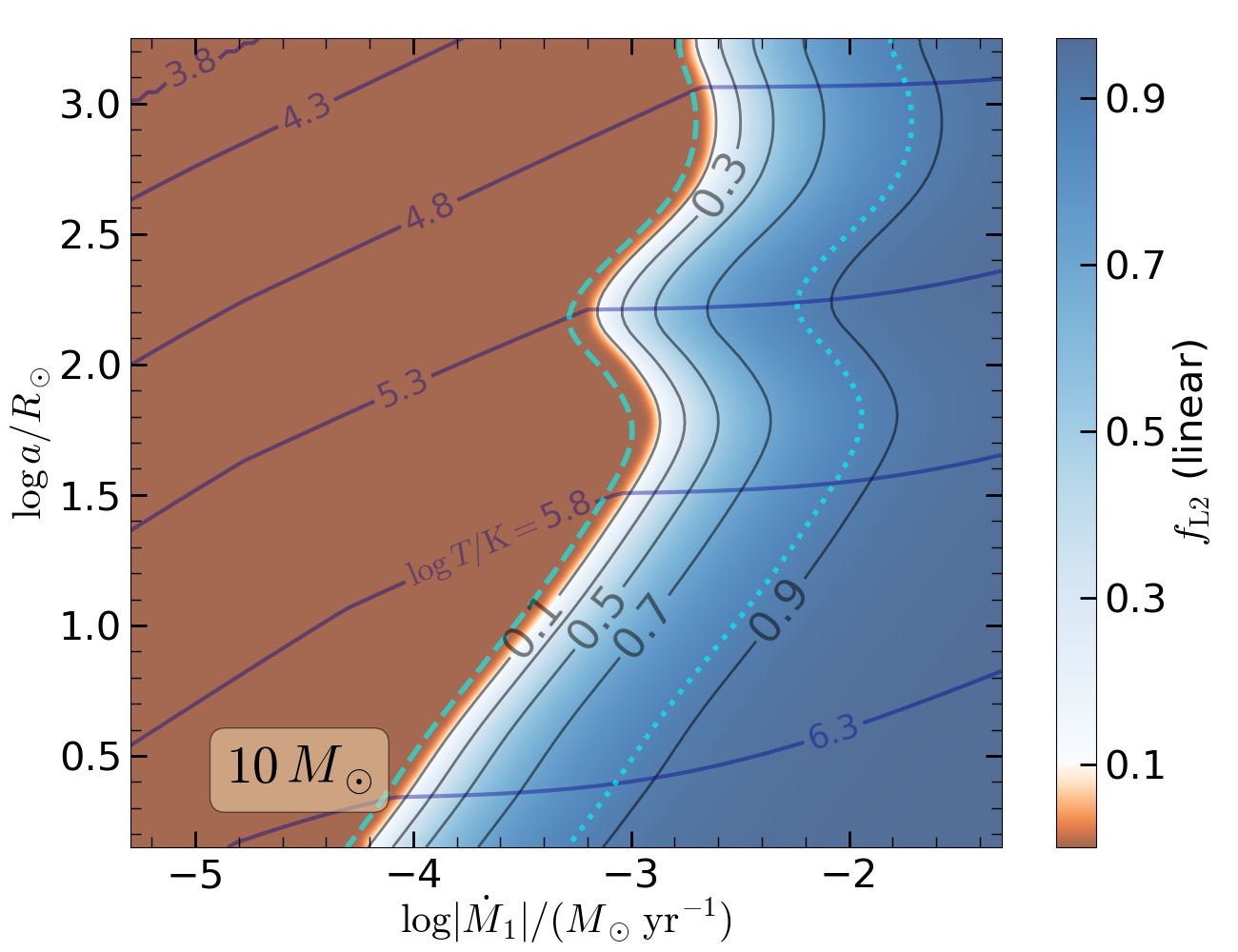}
\includegraphics[width = 0.49\textwidth]{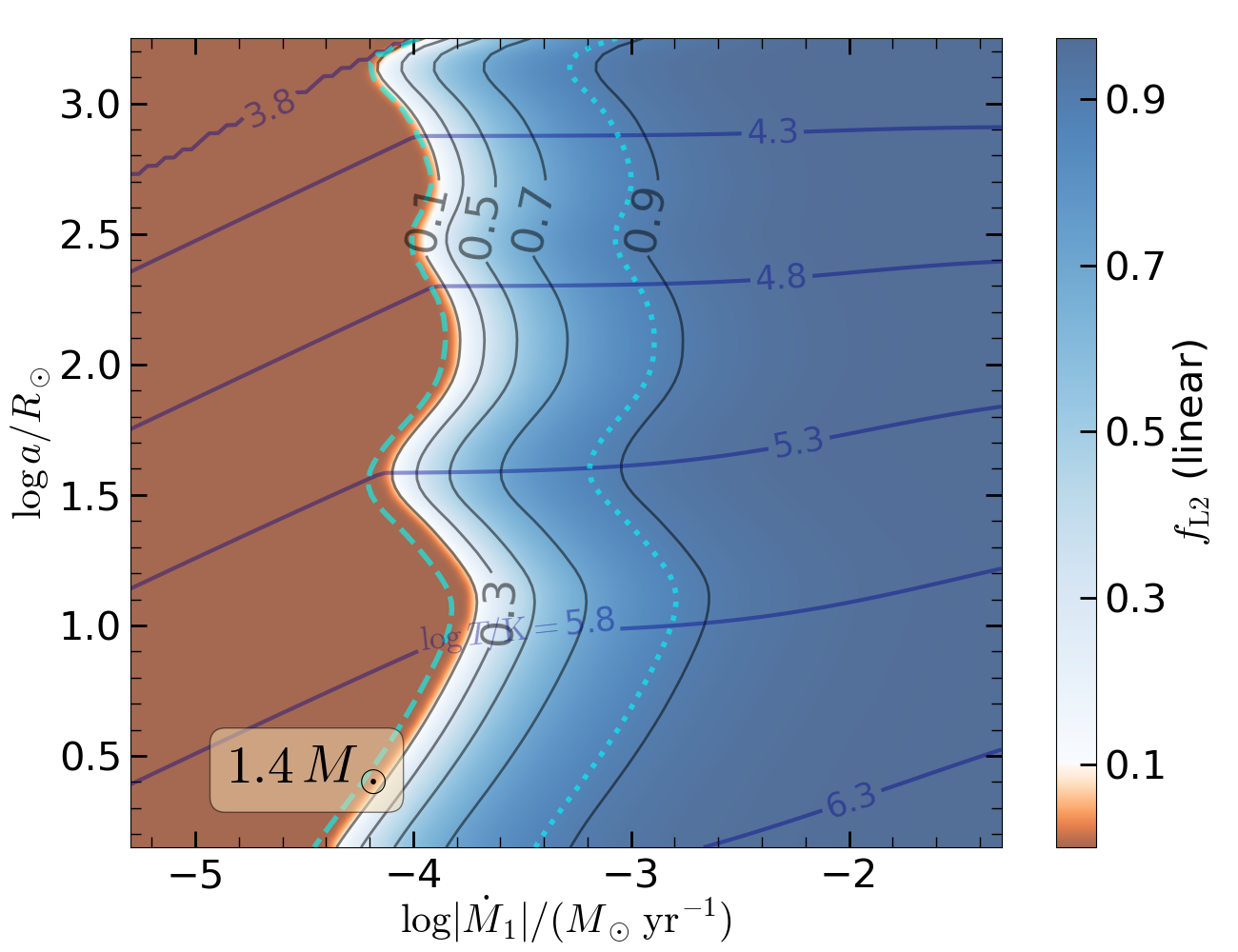}
\caption{The outer disk solutions for a wide range of mass transfer rates $|\dot{M}_1|$ and semimajor axes $a$. The dark blue contours show the disk temperature $T$, and the color-shading and black contours show the L2 mass loss fraction from the outer disk $\fL2^{\rm outer}$ (in linear scale). The left panel is for secondary mass $M_2 = 10\Msun$ (massive star or BH), and the right panel is for $M_2 = 1.4\Msun$ (NS). For both panels, we have fixed the mass ratio $q = M_2/M_1 = 0.5$ and viscosity parameter $\alpha=0.1$. The gas composition is H-rich at solar metallicity ($X=0.7, Z=0.02$). The cyan dashed and dotted lines in each panel indicate where $(Q_{\rm adv} + Q_{\rm L2})/Q_{\rm rad} = 0.1$ and 1, respectively (the latter roughly corresponds to where $GM_2|\dot{M}_1|/\Rd\simeq L_{\rm Edd,2}$). When the local viscous heating rate near $\Rd$ is comparable to the Eddington luminosity of the secondary, the disk becomes geometrically thick, and hence cooling due to the advective ($Q_{\rm adv}$, eq. \ref{eq:Qadv}) and L2-loss ($Q_{\rm L2}$, eq. \ref{eq:QL2}) terms become important. We find that when the mass-transfer rate exceeds a threshold of about $10^{-4}\Msun\rm\, yr^{-1}$ (which is strongly modulated by the opacity variations), a significant fraction of the transferred mass is lost through the L2 nozzle.
}\label{fig:fL2outer}
\end{figure*}

With the above heating and cooling rates, the energy conservation equation (\ref{eq:21}) can be written in the following dimensionless form
\begin{equation}
  \label{eq:27}
\begin{split}
   {9\over 4} - {3U\over 2P}
     \theta^2 &+ {\Phi_{\rm L1}- \fL2 \Phi_{\rm L2} \over  (1-\fL2) GM_2/\Rd} \\
 &= {\Phi(\Rd)\over GM_2/\Rd} + {1\over 2} + {c_3T^4 \over (1-\fL2)^2\kappa } \theta^2,
\end{split}
\end{equation}
where
\begin{equation}
    c_3 = {8\pi^2 a_{\rm r}\alpha c \Rd^2
   \over |\dot{M}_1|^2 \OmgK}. 
\end{equation}
This equation has many interesting features. When $f_{\rm L2}=0$ and ignoring the shock heating terms, we have recovered the thermodynamic equilibrium of a simple viscously accreting disk \citep[e.g.,][]{yuan14_accretion_flow_review}. At very high accretion rates, the diffusion time becomes longer than the viscous time such that the disk is very thick $\theta\sim 1$ (shock heating also slightly increases the thickness). When L2 mass loss is significant ($\fL2\sim 1$), it efficiently cools the disk to a thinner one. Note that in the extreme limit $1-\fL2\ll 1$, the disk might become optically thin, and then the radiative diffusion term on the right hand side needs to be multiplied by a factor of $(1+1/\tau)^{-1}$, which gets rid of the diverging behavior of $(1-\fL2)^{-2}$. We have checked that our solutions have $\tau\gg 1$ throughout the parameter space, so this correction factor is not needed.

Now we have two equations (\ref{eq:43}) and (\ref{eq:27}) for three unknowns: $\theta$, $T$, and $\fL2$. To close the system of equations, we must introduce a model for $\fL2$, the fraction of the transferred mass that is lost through the L2 nozzle. We take the minimum $\fL2$ such that the remaining gas in the disk is not able to reach the L2 potential surface. This solution is a stable equilibrium for the following reason. A larger $\fL2$ than the minimum value cools the disk even more such that there would be a gap between the energy of the mass lost through L2 nozzle and the remaining disk material. This is not the most energetically efficient solution and is disfavored by our physical intuition. On the other hand, a smaller $\fL2$ than the minimum means that the remaining disk material still has enough energy to climb up the potential to the L2 point (provided that radiative cooling can be ignored). This would lead to more mass loss through L2.

In realistic systems, there is likely a transient phase where $\fL2$ can be either higher or lower than its minimum value, and subsequently the system will evolve towards the equilibrium when causal contact between the edge of the disk and the L2 point is established. The equilibrium $\fL2$ is calculated as follows.

In the adiabatic limit, the Bernoulli number of the disk material at radius $\Rd$ is given by
\begin{equation}
  \label{eq:30}
  \mr{Be}(\Rd) = {GM_2/\Rd\over 2} +
    \Phi(\Rd) + h,
\end{equation}
where the specific enthalpy is given by
\begin{equation}
    h = {U + P\over \rho} = {5\kB T\over 2\mu_{\rm g}\mp} +
    {4a_{\rm r} T^4 \over 3\rho}.
\end{equation}
The Bernoulli number describes the total (kinetic plus potential) energy content of the gas and stays constant along a given stream line in a non-viscous and non-turbulent system. It is likely that the shear motion for the gas flows at radii larger than $\Rd$ do not have enough time to develop strong (MHD) turbulence to affect the bulk motion. Thus, disk gas with $\mr{Be}(\Rd)\gtrsim \Phi_{\rm L2}$ is able to reach the L2 potential and hence escape to much larger radii, if we ignore the radiative cooling along the way.

To capture the detailed dynamics of the gas escaping from the disk through the L2 nozzle, one has to carry out radiation-MHD simulations in three dimensions. This is left to be explored in future work. For an analytical estimate, we adopt the following prescription
\begin{equation}
  \label{eq:53}
  \begin{cases}
    \fL2 = 0,\ \ &\mr{if}\ \mr{Be}(\fL2=0)<\Phi_{\rm L2},\\
    \mr{Be}(\fL2) = \Phi_{\rm L2}, \ \ &\mr{otherwise},
  \end{cases}
\end{equation}
This limits the Bernoulli number to less than $\Phi_{\rm L2}$ \citep[see][for a similar treatment but in a different context]{margalit16_Bernoulli_limited_disk}. The maximum disk thickness $\theta_{\rm max} R_{\rm d}$ is given by $\mr{Be}(\fL2) = \Phi_{\rm L2}$,
\begin{equation}
  \label{eq:51}
  {4c_1T^4\over 1-\fL2}\theta_{\rm max}^3 + {5\over 2} c_2 T =
  {\Phi_{\rm L2} - \Phi(\Rd)\over GM_2/\Rd} - {1\over 2},
\end{equation}
where $c_1$ and $c_2$ have been defined in equation (\ref{eq:c1c2}). The above equation can be combined with equation (\ref{eq:43}) to yield
\begin{equation}
  \label{eq:52}
  \theta_{\rm max}^2 = {3c_2T\over 8} + {1\over 4} {\Phi_{\rm L2} - \Phi(\Rd)
    \over GM_2/\Rd} - {1\over 8}.
\end{equation}
If the solution under no L2 mass loss, $\theta(\fL2=0)$, exceeds this maximum thickness $\theta_{\rm max}$, then we require some finite $0<\fL2<1$ so as to maintain $\theta(\fL2) = \theta_{\rm max}$.

The above model allows us to solve for the L2 mass-loss fraction $f_{\rm L2}$ as a function of the mass-transfer rate $\dot{M}_1$ and semimajor axis $a$. In the following, we denote the L2 mass loss fraction from the outer disk near $\Rd$ as $\fL2^{\rm outer}$, so as to differentiate it from the potential L2 mass loss from the inner disk (see \S \ref{sec:inner-disk}). The results from our model, for the accretion disk near radius $\Rd$, are shown in Fig. \ref{fig:fL2outer}, for two different secondary masses $M_2 = 10M_\odot$ (massive star or BH) and $1.4M_\odot$ (NS).  We find that, when the mass-transfer rate exceeds a few times $10^{-4}\Msun\rm\, yr^{-1}$ such that the local viscous heating rate near $\Rd$ becomes comparable to the Eddington luminosity of the secondary $L_{\rm Edd,2}=4\pi GM_2c/\kappa$ (where $\kappa$ is the gas opacity at $\Rd$), an order-unity fraction of the transferred mass is directly lost from the outer disk through the L2 nozzle. 




\subsection{L2 Mass Loss from the Inner Accretion Disk}\label{sec:inner-disk}
In this section, we consider lower mass-transfer rates at which the outer disk is not capable of driving L2 mass loss because the gas is radiating efficiently.
At lower mass-transfer rates for which $\fL2^{\rm outer}\approx 0$, the inner disk can still launch an outflow near the spherization radius $\Rsph\ll \Rd$ where $GM_2|\dot{M}_1|/\Rsph$ is close to the Eddington luminosity of the secondary $L_{\rm Edd,2}$ and hence the disk becomes geometrically thick \citep{shakura73_alpha_disk, begelman78_trapping_radius}, provided that the secondary object is sufficiently small $R_2 < \Rsph$. The spherization radius is given by
\begin{equation}
  \label{eq:54}
  \Rsph = \mr{min}\left[\Rd, {(1-\fL2^{\rm outer})|\dot{M}_1|\kappa\over 4\pi
      c}\right]. 
\end{equation}
At small radii $\Rsph\ll \Rd$ from the secondary, most of the disk material cannot find a (fine-tuned) stream line that directly connects to the L2 nozzle, so the natural way to remove the excessive energy from the system is to launch a wind\footnote{There was a historic debate whether advection-dominated accretion flows (ADAFs) lose a substantial fraction of mass in unbound outflows \citep{narayan94_ADAF, narayan95_ADAF, narayan97_1D_ADAF_no_wind, blandford99_ADIOS, blandford04_2D_ADIOS} or nearly all the material stays bound \citep{stone99_ADAF, abramowicz00_ADAF_no_winds, narayan00_CDAF, quataert00_CDAF, igumenshchev03_3D_ADAF}. Recent large-scale, long-duration numerical simulations of adiabatic accretion flows \citep{narayan12_ADAF, yuan12_rin_PL_index} show that unbound outflows are generated in the regions beyond tens of gravitational radii ($\Rg=GM_2/c^2$) from a black hole accretor, provided that the simulation reaches inflow equilibrium up to about $100\Rg$. In our current case, since photons are largely trapped by the accretion flow at small radii from the accretor $R< \Rsph$, this region of the disk is similar to ADAFs that are radiatively inefficient at very low accretion rates. } with asymptotic speed of the order $\sqrt{GM_2/\Rsph}$ \citep{blandford99_ADIOS}. Therefore, the energy distribution of the gas near $\Rsph$ is in between $-GM_2/2\Rsph$ (most bound) and $+GM_2/2\Rsph$ (most unbound).  Most of the unbound gas is not affected by the binary orbit and quickly escape the system as a fast wind. However, a small fraction of the gas with energy of the order $|\Phi_{\rm L2}|$ can be captured by the binary potential and then flow out of the system through the L2 nozzle. 

Based on the above argument, we propose the L2 mass loss fraction from the inner disk to be
\begin{equation}\label{eq:fL2-inner}
    \fL2^{\rm inner} \simeq (1-\fL2^{\rm outer}) {|\Phi_{\rm L2}| \over GM_2/\Rsph},
\end{equation}
which is based on the assumption that the energy distribution near $\Rsph$ is flat between $\pm GM_2/2\Rsph$.  Since $|\Phi_{\rm L2}|\sim 0.1 GM_2/\Rd$, we see that $\fL2^{\rm inner}$ is of the order $0.1\Rsph/\Rd$ based on our prescription in eq. (\ref{eq:fL2-inner}).


Note that here $L_{\rm Edd,2}$ is based on the gas opacity near $\Rsph$ and this is generally different from that near $\Rd$. The gas temperature and  density near $\Rsph$ can be estimated (for disk thickness $\theta\simeq 1$) by
\begin{equation}
    a_{\rm r}T^4 + {3\rho \kB T\over 2\mu_{\rm g} \mp }\simeq {GM_2 \rho\over \Rsph},\ 
    \rho \simeq {(1-\fL2^{\rm outer})|\dot{M}_1| \over 2\pi \alpha \Omega_{\rm K}(\Rsph) \Rsph^3}.
\end{equation}
We note that, for outflows to be launched from the inner disk, the secondary object must be sufficiently small $R_2<\Rsph$. Otherwise $\fL2^{\rm inner}=0$, and in this case the secondary gains mass at a rate $(1-\fL2^{\rm outer})|\dot{M}_1|$ until it reaches near the break-up rotation rate.

\begin{figure}
  \centering
\includegraphics[width = 0.49\textwidth]{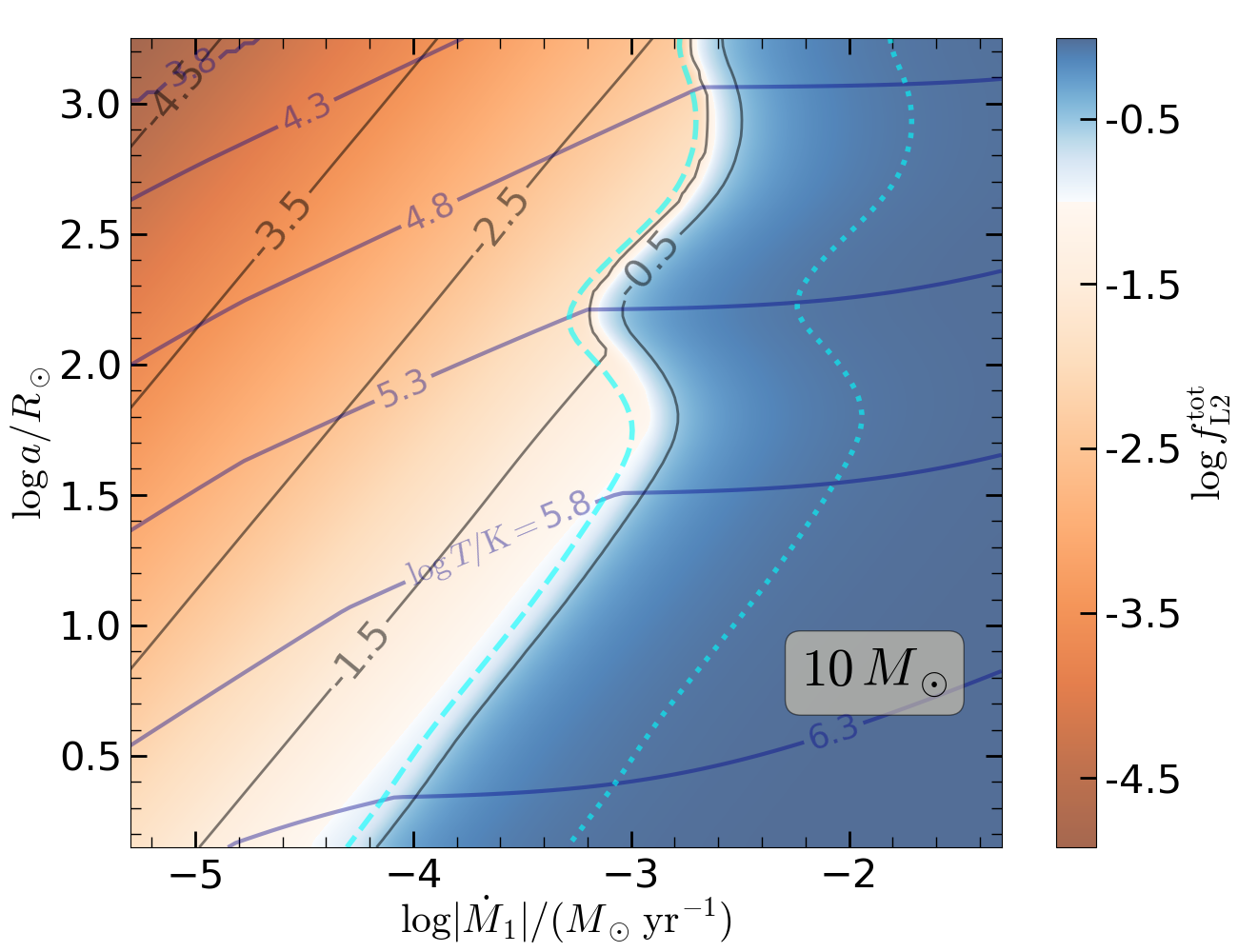}
\includegraphics[width = 0.49\textwidth]{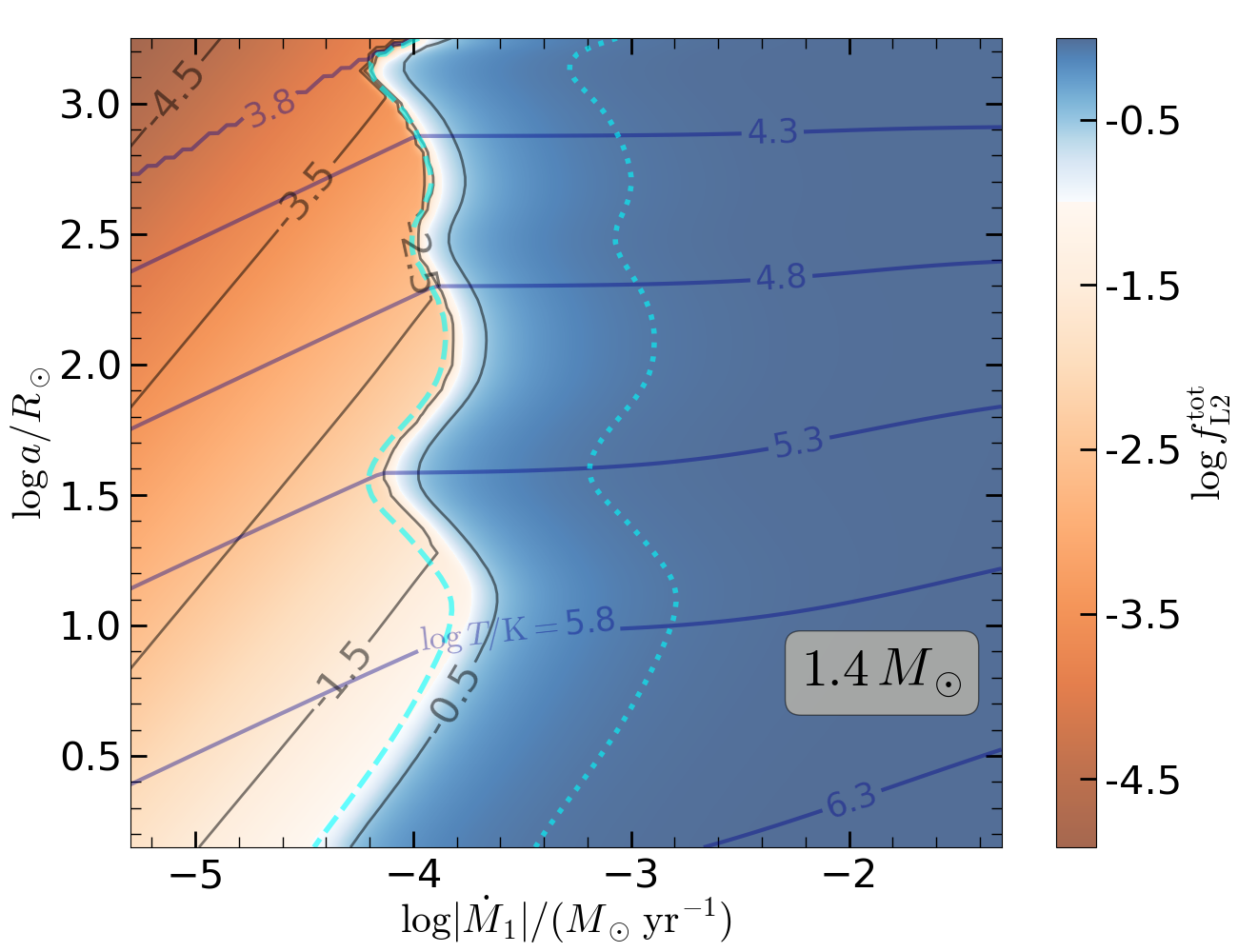}
\caption{The total L2 mass loss fraction (eq. \ref{eq:fL2-total}) from the binary system for different mass transfer rates $|\dot{M}_1|$ and semimajor axes $a$, including contributions from the inner ($\fL2^{\rm inner}$) and outer ($\fL2^{\rm outer}$) disk. Here we assume that the accretor is a compact object. The color-shading and black contours show $\mr{log}_{10}\fL2$, and the dark blue contours show the gas temperature in the outer disk near $\Rd$. The upper panel is for secondary mass $M_2 = 10\Msun$ (BH), and the lower panel is for $M_2 = 1.4\Msun$ (NS). For both panels, we have fixed the mass ratio $q = M_2/M_1 = 0.5$ and viscosity parameter $\alpha=0.1$. The gas composition is H-rich at solar metallicity ($X=0.7, Z=0.02$). The cyan dashes and dotted lines in each panel indicate where $(Q_{\rm adv} + Q_{\rm L2})/Q_{\rm rad} = 0.1$ and 1, respectively.
}\label{fig:fL2tot}
\end{figure}

Finally, we combine the L2 mass loss from the outer and inner disk to obtain the \textit{total} L2 mass loss fraction
\begin{equation}\label{eq:fL2-total}
    \fL2 = \fL2^{\rm outer} + \fL2^{\rm inner}.
\end{equation}
This is shown in Fig. \ref{fig:fL2tot}, for two different secondary masses $M_2 = 10M_\odot$ and $1.4M_\odot$, under the assumption that the secondary is a compact object with $R_2\ll \Rsph$ (a BH or NS).

We find that the contribution to L2 mass loss from the inner disk is small: $\fL2^{\rm inner}$ is at most a few percent in the parameter space where $\fL2^{\rm outer}$ is negligible. The overall trend is that, at a fixed binary separation, the total L2 mass loss fraction increases linearly (since $\Rsph\propto |\dot{M}_1|$) with the mass-transfer rate until the viscous heating rate in the outer disk becomes comparable to $L_{\rm Edd,2}$ and, beyond this mass-transfer rate, nearly all the transferred mass is lost through the L2 nozzle. \note{The orbital evolution of the system is only strongly affected by the angular momentum loss associated with L2 mass loss when $\fL2$ becomes close to order unity (at which point $\fL2\approx \fL2^{\rm outer}$), so it is appropriate to ignore the effects of a small $\fL2^{\rm inner}$ on the orbital evolution.}

\section{Binary Orbital Evolution and Accretion onto the Secondary}\label{sec:orbital_evolution}
In this section, we consider the effects of L2 mass loss as well as accretion onto the secondary on the binary orbital evolution. The key result is that the angular momentum loss associated with the CBO tends to cause the orbit to shrink faster or expand less than in the case without L2 mass loss \citep[e.g.,][]{vandenheuvel73_orbital_evolution}.

\begin{figure}
  \centering
\includegraphics[width = 0.49\textwidth]{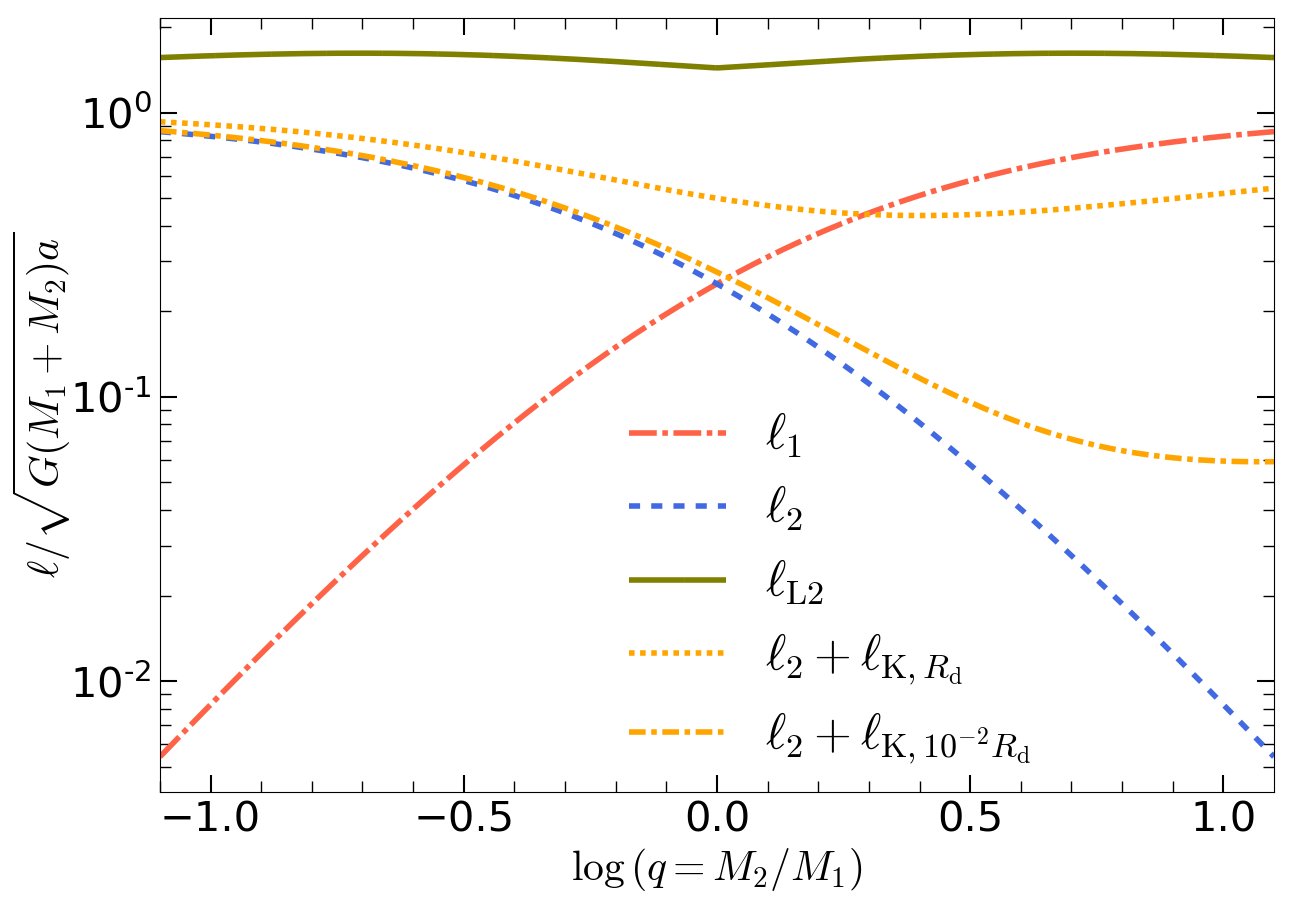}
\caption{The specific angular momenta $\ell_1$ (donor's orbital motion, red long-dash-dotted line), $\ell_2$ (accretor's orbital motion, blue dashed line), $\ell_{\rm L2}$ (L2 point at corotation, greed solid line), $\ell_2 + \ell_{\mr{K}}(R_{\rm d})$ (accretor's orbital motion plus disk Keplerian rotation at the circularization radius $\Rd$, orange dotted line), $\ell_2 + \ell_{\mr{K}}(10^{-2}R_{\rm d})$ (accretor's orbital motion plus disk Keplerian rotation at a distance $10^{-2}\Rd$ from the accretor, orange short-dash-dotted line), all in units of the specific orbital angular momentum per unit reduced-mass $\sqrt{G(M_1+M_2)a}=\Omega a^2$. This shows that L2 mass loss generally carries much more angular momentum per unit mass away from the binary system than other channels such as fast stellar wind (red/blue lines) or disk wind (orange lines).
}\label{fig:angmom}
\end{figure}

\begin{figure}
  \centering
\includegraphics[width = 0.5\textwidth]{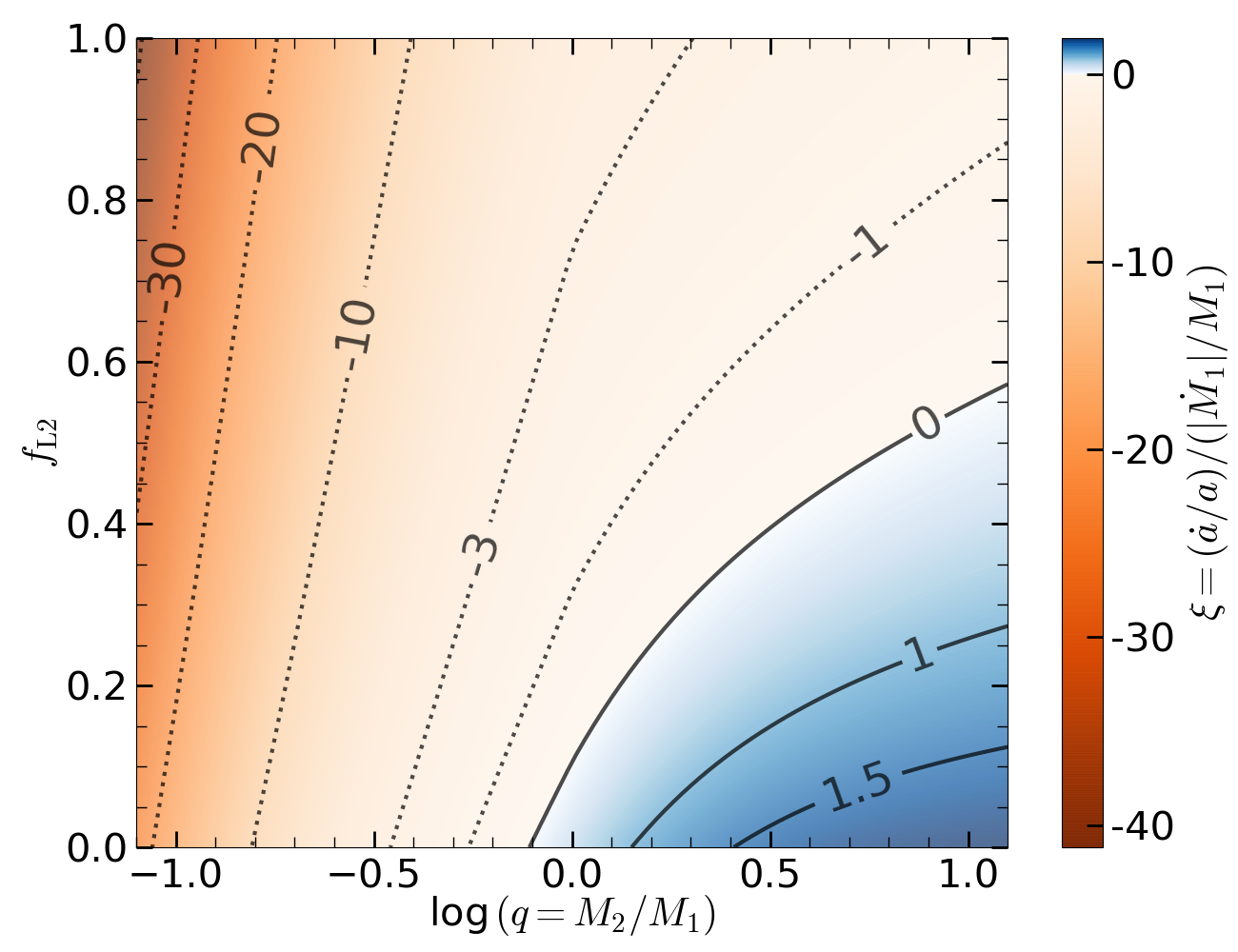}
\caption{The ratio between the dimensionless orbital shrinking rate and the mass loss rate,  $\xi = (\dot{a}/a)/(|\dot{M}_1|/M_1)$, as a function of binary mass ratio $q$ and the total L2 mass loss fraction $\fL2$, according to eq. (\ref{eq:57}). The binary orbit shrinks (or expands) when $\xi <0$ (or $\xi>0$). It is assumed that the L2 mass loss carries the corotational angular momentum at the L2 point (taking $g_{\rm b}=1$ in eq. \ref{eq:L2_AM}), and the disk-rotational component of the wind angular momentum is ignored (taking the limit of $\Rsph\ll a$). The secondary is taken to be a compact object with negligible accretion rate ($\dot{M}_2= 0$). We see that for $q\lesssim 1/3$ (the donor is much more massive), the orbit rapidly shrinks on a timescale that is many times shorter than the mass-loss timescale $M_1/|\dot{M}_1|$ of the primary. For $q\gtrsim 1$ (the donor is less massive), the orbit may shrink as a result of significant L2 mass loss, as opposed to the usual expectation of orbital expansion under $\fL2=0$. 
}\label{fig:orbital_evolv}
\end{figure}

In the center of mass frame, the specific angular momentum of the gas corotating at the L2 point is given by
\begin{equation}
  \label{eq:33}
  \ell_{\rm L2} = \Omega a^2(x_{\rm L2}-\mu)^2.
\end{equation}
In the following, we assume that the asymptotic angular momentum of the gas escaping from the L2 nozzle is $g_{\rm b} \ell_{\rm L2}$, where $g_{\rm b}$ is a factor of order unity describing the angular momentum gain due to the torque from binary. Hydrodynamic simulations by \citet[][their Fig. 10]{pejcha16_L2_stream_shape} showed that the angular momentum gain is rather modest $g_{\rm b}\in (1, 1.3)$. If the gas flowing out from the L2 point stays bound to the binary and forms an ``decretion'' disk, then the long-term viscous evolution of the disk may further extract angular momentum from the binary and hence may lead to a higher effective $g_{\rm b}$ that is not captured by the simulations by \citet{pejcha16_L2_stream_shape}. On the other hand, at extremely high mass-transfer rates such that $\fL2\sim 1$, the L2 mass loss may not achieve corotation with the orbit at the L2 point \citep[see][]{macleod18_L2_loss_binary_merger}, so we may expect $g_{\rm b} < 1$ in such cases. Detailed numerical simulations would be required to carefully determine $g_{\rm b}$.

When a fraction $f_{\rm L2} \approx \fL2^{\rm outer}$ (ignoring $\fL2^{\rm inner}$) of the transferred mass $\dot{M}_1$ is lost through the L2 nozzle, the angular momentum loss rate from the binary is given by
\begin{equation}
  \label{eq:L2_AM}
  \dot{J}_{\rm L2} = -f_{\rm L2}|\dot{M_1}| g_{\rm b}\ell_{\rm L2}.
\end{equation}
As we show later (eq. \ref{eq:57}), in the limit $\fL2\approx 1$, the orbital separation shrinks as long as $g_{\rm b}$ exceeds a critical $g_{\rm b,c} = 0.5\mu(1+\mu)/(x_{\rm L2}-\mu)^2$, which is in the range $0.1\lesssim g_{\rm b,c}\lesssim 0.4$ for $0.3<q<3$. Thus, the general effect of $\dot{J}_{\rm L2}$ is to push the two stars closer to each other.

When there is a large dynamical range between the surface or innermost stable circular orbit  (ISCO) of the secondary $R_2$ and the spherization radius $\Rsph$, then the accretion disk is expected to take a self-similar or power-law \citep{blandford99_ADIOS} radial profile of mass accretion rate in the radius range $R_2 < R < \Rsph$, 
\begin{equation}
  \label{eq:accretion_rate}
\begin{split}
  \dot{M}_{\rm acc}(R) &\simeq (R/\Rsph)^p (1-f_{\rm L2})|\dot{M}_1|,
\end{split}
\end{equation}
where the power-law index $p$ is in general between 0 (no wind) and 1 (the maximum allowed by energy conservation). Numerical simulations of hot accretion flows typically find $p\in (0.3, 0.8)$, but this is still rather uncertain \citep[see][and references therein]{yuan14_accretion_flow_review}. The basic result is that, if the size of the secondary is small $R_2\ll \Rsph$ (e.g., in the case of a compact object), then only a very small fraction of the transferred mass is actually accreted onto the secondary,
\begin{equation}
  \label{eq:39}
  \dot{M}_2 \equiv \dot{M}_{\rm acc}(R_2) = \mr{min}\left[(R_2/R_{\rm sph})^p, 1\right] (1-\fL2)|\dot{M}_1|.
\end{equation}
We assume that the wind directly escapes from the system in a quasi-spherical manner due to its fast speed (ignoring the small fraction that is re-captured by the binary's potential). \note{The angular momentum carried away by the wind has two components due to the Keplerian rotation at the disk surface and orbital motion of the secondary. The specific angular momentum of the wind, averaged over the total wind mass-loss rate of $(1-f_{\rm L2}) |\dot{M}_1| - \dot{M}_2$, is given by an integral over the Keplerian angular momentum profile $\ell_{\rm K}(R) = \sqrt{GM_2R}$ over the entire disk,
\begin{equation}\label{eq:wind_AM}
\begin{split}
    \ell_{\rm w} &= {\int_{R_2}^{\Rsph} \d R\, (\d \dot{M}_{\rm acc}/ \d R) \sqrt{GM_2R} \over (1-f_{\rm L2}) |\dot{M}_1| - \dot{M_2}} + \ell_2\\
    &= {2p\over 2p+1} {1 - (R_2/\Rsph)^{p+1/2} \over 1 - (R_2/\Rsph)^{p}} \sqrt{GM_2\Rsph} + \ell_2,
\end{split}
\end{equation}
where the accretion rate profile $\dot{M}_{\rm acc}(R)$ is given by eq. (\ref{eq:accretion_rate}) and
\begin{equation}
    \ell_2 = \Omega a^2 (1-\mu)^2 = (1+q)^{-2} \sqrt{G(M_1 + M_2)a}
\end{equation}
is the specific angular momentum of the secondary. The accretion onto the secondary also gives an angular momentum loss rate of $\dot{M}_2 \sqrt{GM_2R_2}$, which is usually negligible unless the accretion rate is low or the secondary radius is large such that $R_2\sim \Rsph$. The ``isotropic re-emission'' prescription often adopted in binary evolution studies \citep{paxton15_mesa_binary} corresponds to an assumption of $\ell_{\rm w}\approx \ell_2$, which is only applicable if $\sqrt{q(1+q)^3\Rsph/a}\ll 1$. We caution that this assumption is violated if $\Rsph$ is a sizable fraction of $\Rd$ (achieved at high mass-transfer rates) and $q\gtrsim 2$. Various specific angular momenta are shown in Fig. \ref{fig:angmom}.}



In the following, we ignore the small angular momentum contributions by the spins of the two stars.
The total angular momentum loss rate of the binary system is
\begin{equation}
  \label{eq:55}
  \dot{J} = \dot{J}_{\rm L2} + \dot{J}_{\rm w},
\end{equation}
where
\begin{equation}
  \label{eq:37}
  \dot{J}_{\rm w} = -\left[(1-f_{\rm L2}) |\dot{M}_1| - \dot{M}_2\right]\ell_{\rm w}.
\end{equation}
Since the total mass loss rate is $\dot{M} = \dot{M}_1 + \dot{M}_2$, we make use of the expression for total angular momentum of a Keplerian circular orbit $J =\Omega a^2M_1 M_2/M$ ($M$ being the total mass), and obtain the orbital shrinking (or expanding) rate
\begin{equation}
  \label{eq:38}
  {\dot{a}\over a} = 2{\dot{J}\over J} - 2{\dot{M}_1 \over M_1} -
  2{\dot{M}_2 \over M_2} + {\dot{M} \over M}.
\end{equation}

In the limit where the secondary is a compact object, $(R_2/\Rsph)^p\ll 1$, $\dot{M}_2\ll |\dot{M}_1|$, and $\dot{M} \approx \dot{M}_1$, we \note{obtain
\begin{dmath}
  \label{eq:56}
  {\dot{J}\over J}\approx {\dot{M}_1\over \mu M_1} \left\{g_{\rm b}\fL2(x_{\rm L2} - \mu)^2 +\\
    (1-\fL2)\lrsb{(1-\mu)^2 + {2p\over 2p+1} \sqrt{\mu \Rsph\over a} } \right\},
\end{dmath}
and hence
\begin{dmath}
  \label{eq:57}
  \xi = {\dot{a} M_1\over a \dot{M}_1} \approx - (1+\mu) + {2\over \mu} \left\{g_{\rm b}\fL2(x_{\rm L2} - \mu)^2 +\\
    (1-\fL2)\lrsb{(1-\mu)^2 + {2p\over 2p+1} \sqrt{\mu \Rsph\over a} } \right\}.
\end{dmath}
In Fig. \ref{fig:orbital_evolv}, we show the ratio between the dimensionless orbital shrinking rate $\dot{a}/a$ and the mass-loss rate $\dot{M}_1/M_1$ as a function of mass ratio $q$ and $f_{\rm L2}$, for $g_{\rm b}=1$ (the L2 mass loss carrying the corotational angular momentum of the L2 point). Here we ignore the disk-rotational angular momentum component of the wind by imposing $\Rsph\ll a$. The qualitative effects of $\fL2$ are unaffected by the choice of $\Rsph/a$.}

We find that, if $q\lesssim 1/3$ and $f_{\rm L2}\simeq 1$, the orbit rapidly shrinks on a timescale that is a few to 10 percent of the mass loss timescale of the primary, $M_1/|\dot{M}_1|$. Moreover, at sufficiently high mass-transfer rates such that $f_{\rm L2}\gtrsim 0.5$, the orbit always shrinks for practically all mass ratios. \note{Note that, under the (in our opinion questionable) assumptions of $f_{\rm L2}=0$ and $\Rsph/a= 0$}, our eqs. (\ref{eq:56}) and (\ref{eq:57}) reduce to the conventional picture  where the orbit starts to expand when $q>0.781$ \citep{postnov14_binary_evolution}. Faster orbital shrinking due to L2 mass loss ($\fL2\neq 0$) has implications for the formation of compact binary gravitational wave sources, which will be discussed in a companion paper.

\section{IR Excess from Circum-Binary Outflow}\label{sec:EM_signatures}

\begin{figure*}
  \centering
\includegraphics[width = 0.7\textwidth]{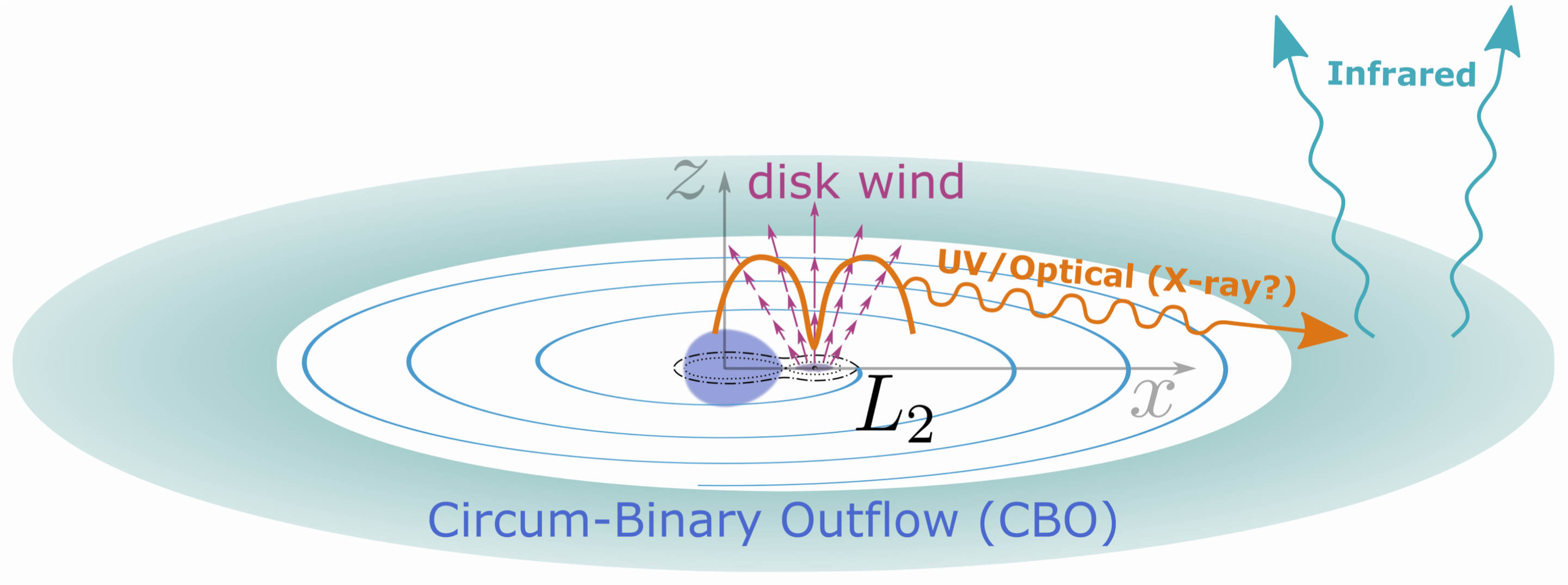}
\caption{Schematic picture of a binary system undergoing mass transfer at a high rate greater than $10^{-4}\,\msunyr$ such that a significant fraction of the transferred mass is lost through the L2 point, forming a circum-binary outflow (CBO). Immediately after flowing out from the L2 nozzle, the gas is in the form of a spiral-shaped supersonic cold stream. At large distances (5 to 10 times the semi-major axis) from the binary, the gas undergoes internal shocks and forms an axisymmetric geometrically thin outflow. The inner accretion disk near the accretor launches a fast wind whose scattering photosphere is indicated by a thick orange curve. The UV/optical (and a small flux of X-ray) photons from the hot photosphere irradiate the CBO, which reprocesses the incident radiation into longer wavelength photons in the infrared (IR) bands. This model explains the IR excess observed in some ULXes \citep[e.g.,][]{lau19_ULX_MIR}.
}\label{fig:CBO}
\end{figure*}

In this section, we discuss the radiative appearance of the L2 mass loss, which is assumed to be in the form of a circum-binary outflow\footnote{For a wide range of mass ratios $0.06\lesssim q\mbox{ (or $q^{-1}$)}\lesssim 0.8$, cold gas streaming out from the L2 nozzle gains energy from the binary's tidal torque and becomes unbound \citep{shu79_L2_stream, pejcha16_L2_stream_shape}.} (CBO). The physical picture is that the inner disk drives a quasi-spherical fast wind and the UV radiation escaping from the wind photosphere is reprocessed into IR emission by the CBO. A schematic picture is shown in Fig. \ref{fig:CBO}.

Let us consider the case that a fraction $\fL2\approx \fL2^{\rm outer}$ of the mass transfer-rate $|\dot{M}_1|$ is channeled into the CBO and the rest $(1-\fL2)$ is driven away as a quasi-spherical wind. The scattering photosphere of the wind is located at radius
\begin{equation}\label{eq:Rsca}
\begin{split}
    \Rs &= {(1-\fL2)|\dot{M}_1| \kaps \over 4\pi \vw} \\
    &= 0.4\mr{\,AU}\, {\kaps\over 0.34\rm\, cm^2\,g^{-1}} {(1-\fL2)|\dot{M}_1| \over 10^{-3}\,\msunyr} {0.01c\over \vw},
\end{split}
\end{equation}
where we have taken a scattering opacity $\kaps\equiv 0.34\rm\, cm^2\,g^{-1}$ for H-rich composition and the wind velocity $\vw$ is set by the spherization radius \citep[where most of the disk outflow originates,][]{begelman06_superEddington_wind}
\begin{equation}\label{eq:vw}
    \vw\sim \sqrt{GM_2\over \Rsph}= \sqrt{L_{\rm Edd,2}\over (1-\fL2)|\dot{M}_1|}.
\end{equation}
For high mass accretion rates $(1-\fL2)|\dot{M}_1|\sim 10^{-3}\,\msunyr$, the wind velocity is about $10^{-2}\rm\, c$ and the wind photospheric radius can be comparable to or larger than the orbital separation. This makes it difficult to observe the donor star.

The luminosity and spectrum of the radiation escaping from the wind depends on the nature of the accretor as well as the viewing angle. In the case of a compact-object accretor (BH or NS), the bolometric luminosity of the entire system is dominated by the innermost regions of the accretion disk. Only a small fraction of the transferred mass reaches near the ISCO (eq. \ref{eq:accretion_rate}), so photons decouple from the gas within ten times the ISCO radii even though the wind velocity is close to the speed of light \citep{jiang14_rad_MHD, jiang19_rad_MHD_superEddington}. From viewing angles not far from the rotational axis of the disk, the system is expected to appear as a bright ULX source with X-ray luminosity $L_{\rm X}> 10^{39}\rm\, erg\,s^{-1}$ \citep{kaaret17_ULXs}.
However, due to geometric beaming \citep{narayan17_ULX}, the X-ray photons from the fastest wind do not reach the CBO which is located near the equatorial plane. 

Instead, the irradiation of the CBO is dominated by the photons that are reprocessed by the wind launched from near the spherization radius $\Rsph$ and then escape at the scattering photosphere given by eq. (\ref{eq:Rsca}). Hereafter, we denote the luminosity and color temperature of the radiation escaping from the scattering photosphere of the wind as $L_{\rm s}$ and $T_{\rm s}$. We expect $L_{\rm s}$ to be comparable to (or perhaps slightly higher than) the Eddington luminosity $L_{\rm s}\sim L_{\rm Edd,2}\sim 10^{38}(M_2/\Msun)\rm\, erg\,s^{-1}$, whereas the temperature $T_{\rm s}$ is set by the radius of the last-absorption $R_{\rm th}$, where the effective/thermalization optical depth is unity \citep[][their eq. 1.120]{rybicki86_book}
\begin{equation}
    \tau_{\rm eff} = \sqrt{3\kappa_{\rm a}(\kappa_{\rm a} + \kaps)} \rho_{\rm w} R_{\rm th} \simeq 1,
\end{equation}
where $\kappa_{\rm a}$ is the absorption opacity for photons near the peak of the local spectral energy distribution (SED) and $\kaps$ is the scattering opacity. The opacity depends on the wind density $\rho_{\rm w} = (1-\fL2)|\dot{M}_1|/(4\pi R^2\vw)$ as well as the local intensity/spectrum of the radiation field. In the limit $\kappa_{\rm a}\ll \kappa_{\rm s}$, the Rosseland-mean optical depth ($\sim$ scattering optical depth) at $R_{\rm th}$ is $\tau(R_{\rm th}) \simeq R_{\rm s}/R_{\rm th} \simeq \sqrt{\kaps/3\kapa}$, and the luminosity of the escaping photons is given by
\begin{equation}\label{eq:Ls}
    L_{\rm s} \simeq 4\pi R_{\rm th}^2 {a_{\rm r}T_{\rm s}^4 c\over \tau} \simeq 4\lrb{3\kapa\over \kaps}^{3/2} (4\pi R_{\rm s}^2 \sigma_{\rm SB} T_{\rm s}^4),
\end{equation}
where $\sigma_{\rm SB}=a_{\rm r}c/4$ is the Stefan-Boltzmann constant. Therefore, we obtain a rough estimate of the color temperature of the radiation escaping from the wind
\begin{equation}
\begin{split}
    T_{\rm s} \simeq \,& 2.5\times10^4\mr{\,K}\, \lrb{L_{\rm s}\over 10^{39}\rm\, erg\,s^{-1}}^{1\over 4}
    \lrb{\kapa\over 0.01\rm\, cm^{2}\, g^{-1}}^{-{3\over 8}} \\
    &\lrb{\kaps\over 0.34\rm\, cm^2\,g^{-1}}^{-{1\over 8}} \lrb{(1-\fL2)|\dot{M}_1| \over 10^{-3}\,\msunyr}^{-{1\over 2}} \lrb{\vw \over 0.01c}^{1\over 2}.
\end{split}
\end{equation}
It is beyond the scope of this paper to solve the full frequency-dependent radiative transport problem throughout the wind \citep[see][for general considerations]{shen15_outflow_rad_transfer}. Our preliminary calculations based on Cloudy\footnote{Version 17.01 of the code last described by \citet{Ferland17_Cloudy}.} showed that, for the fiducial parameters above, the bound-free absorption opacity near photon energy $3\kB T_{\rm s}\sim 10\mr{\,eV}\, (T_{\rm s}/3\times10^{4}\rm \,K)$ is of the order $10^{-2}\rm\, cm^2\, g^{-1}$ at $R_{\rm th}$ (where the scattering optical depth is $\sim$3). Since the wind velocity scales as $\vw\propto [(1-\fL2)|\dot{M}_1|]^{-1/2}$ (eq. \ref{eq:vw}) and the absorption opacity $\kapa$ generally decreases with accretion rate, we see that the photospheric temperature is a decreasing function of the mass accretion rate --- bright emission in the optical band is only possible at high accretion rates.

Observationally, a large fraction of ULXes have optical counterparts \citep[e.g.,][]{tao11_opt_counterparts, gladstone13_opt_counterpart, vinokurov18_opt_counterpart} and the brightest ones have spectral luminosity of $\nu L_\nu\sim 10^{38}\rm\, erg\,s^{-1}$ in the $B$-band (e.g., Holmberg II X-1, Holmberg IX X-1, M101 ULX-1). These sources with brightest optical luminosities are consistent with the emission from the aforementioned wind with $L_{\rm s}\sim 10^{39}\rm\, erg\,s^{-1}$ and $T_{\rm s}\simeq 3\times10^4\rm\, K$. Follow-up observations of some of these sources show that the flux density at longer wavelengths (near- and mid-IR) is much higher than the simple power-law extrapolation from the optical bands \citep{heida14_NIR_counterparts, lopez17_NIR_counterparts, lau17_MIR_counterparts, lau19_ULX_MIR}. In the following, we show that the IR excess is consistent with the reprocessed emission from the CBO.

Let us consider the CBO to be a geometrically thin but optically thick sheet near the orbital plane of the binary and that the CBO extends from an inner radius $R_{\rm min}$ to outer radius $R_{\rm max}$. The L2 mass loss flows out in the form of a supersonic thin spiral-shaped stream, which then undergoes internal shocks and forms an axisymmetric outflow at a distance 5 to 10 times the binary separation \citep{pejcha16_shock_powered_transients, pejcha16_L2_stream_shape}. Thus, the inner radius of the reprocessing sheet is at
\begin{equation}
    R_{\rm min}\sim 10a.
\end{equation}
In the test particle limit \citep[as considered by][]{shu79_L2_stream}, the motion of a fluid element initially corotating at the L2 point is controlled by the binary's tidal torque, and it can acquire positive energy with asymptotic speed $0<\veq \lesssim (1/3)\sqrt{GM/a}$ ($M$ being the total mass) provided that the mass ratio is $q\lesssim 0.8$ or $q\gtrsim 1.3$. The importance of pressure effects have been studied by \citet{pejcha16_L2_stream_shape, hubova19_L2_loss_fate}, who found that if the gas leaving the L2 point has significant thermal energy and is radiatively inefficient, then the radial pressure gradient can accelerate the gas to a higher asymptotic speed than in the test particle limit. Having these uncertainties in mind, we take
\begin{equation}\label{eq:veq}
    \veq \simeq {1\over 3}\sqrt{GM\over a} \simeq 46 \mr{\,km\,s^{-1}} \lrb{M\over 10\Msun}^{1\over 2} \lrb{a\over 100\Rsun}^{-{1\over 2}}.
\end{equation}
The vertical extent of the CBO also depends on the thermal energy at the L2 point. In the following, we adopt a fiducial value for the height-to-radius ratio, $\fOmg \equiv H/R = 0.1 f_{\Omega,-1}$. Therefore, the density profile of the CBO is given by
\begin{equation}\label{eq:rhoeq}
    \rhoeq(R) = {\fL2 |\dot{M}_1|\over 4\pi \fOmg R^2 \veq}.
\end{equation}
The outer radius $R_{\rm max}$ is bounded by the age of the CBO ejection $\veq t_{\rm age} \sim 10^{17}\mr{\,cm}\, (\veq/50\mr{\,km\,s^{-1}}) (t_{\rm age}/10^3\mr{\, yr})$.  Another important constraint is the \textit{radial} optical depth for incident photons $\tau_{\rm eq}(R) = \rhoeq \kappa_{\rm P} R$, where $\kappa_{\rm P}$ is the Planck-mean opacity of the photo-ionized region of the CBO (our fiducial opacities below are estimated based on Cloudy simulations). If $\tau_{\rm eq} < 1$, the CBO is no longer an efficient reprocessor, and hence the maximum radial extent is set by $\tau_{\rm eq}(R_{\rm max}) = 1$, i.e.
\begin{equation}\label{eq:Rmax}
\begin{split}
    R_{\rm max} =\,& {\fL2 |\dot{M}_1| \kappa \over 4\pi \fOmg \veq} \simeq 1\times 10^{16}\mr{\,cm}\, {\kappa_{\rm P}\over 1\rm\, cm^2\,g^{-1}}\\
    & {\fL2 |\dot{M}_1|\over 10^{-3}\, \msunyr} {50\mr{\, km\,s^{-1}} \over \veq} {0.1\over \fOmg}.
\end{split}
\end{equation}
At radii $R_{\rm min}<R\ll R_{\rm max}$, only a thin surface layer of the CBO is directly irradiated by the source photons \citep{chiang97_passive_disk}, whereas at much larger radii $R\gg R_{\rm max}$, the entire CBO is photo-ionized by the UV photons from the disk wind. The photo-ionized regions have temperature $T\sim 10^4\rm\, K$ \citep[as given by the balance between photo-electric heating and cooling due to collisional excitation,][]{draine11_ISM_book} and isothermal sound speed $c_{\rm s} \simeq \sqrt{1.5\kB T/\mp}\simeq 10(T/10^4\rm\, K)\, km\,s^{-1}$, so the gas undergoes vertical expansion to reach a height-to-radius ratio of $c_{\rm s}/\veq\sim 0.2$. The vertical expansion reduces the density as compared to that in eq. (\ref{eq:rhoeq}) and hence $R_{\rm max}$ should be self-consistently reduced. More physically, the low-density gas at a few scale-heights away from the mid-plane is heated to higher temperatures close to $T_{\rm s}$ and expands faster in the vertical direction. Some of the vertically extended gas may also interact with the fast disk wind (via a turbulent shearing layer), which may cause the entire CBO to evaporate. In the following, we only consider the region $R\ll R_{\rm max}$, where the bulk of the CBO can be considered as a dynamically cold gas sheet. 




The flux received by the reprocessing sheet at radius $R \in (R_{\rm min}, R_{\rm max})$ is given by the following integral
\begin{equation}\label{eq:reflection_flux}
\begin{split}
    F(R) & = I_{\rm s} R_{\rm s}^2 \int_0^{\theta_{\rm s}} \sin\theta \d \theta \int_{-{\pi\over 2}}^{\pi\over 2} \d \phi {(R\cos\theta - R_{\rm s}) R_{\rm s} \sin\theta \cos\phi \over (R^2 + R_{\rm s}^2 - 2R R_{\rm s} \cos\theta)^2} \\
    &= I_{\rm s} \lrb{R_{\rm s}\over R}^3  \int_0^{\theta_{\rm s}} \d \theta { 2 (\cos\theta - R_{\rm s}/R) \sin^2\theta \over [1 - 2  (R_{\rm s}/R)\cos\theta + R_{\rm s}^2/R^2]^2},
\end{split}
\end{equation}
where $\theta$ and $\phi$ are the polar and azimuthal angles in a spherical coordinate system with the origin at the center of the photosphere, polar axis pointing in the direction of a surface element on the reprocessing sheet, and the sheet lying in the $\phi=\pm \pi/2$ plane. We have also defined $\theta_{\rm s} = \mr{cos}^{-1}(R_{\rm s}/R)$ as the maximum polar angle contributing to the incoming flux and $I_{\rm s} = \int \d \nu I_{\mr{s},\nu}$ as the frequency-integrated intensity at the wind photosphere. The angular dependence of the intensity is taken to be isotropic at the photosphere, whereas in reality this is affected by limb darkening (which would change the results at an order-unity level). 

For simplicity, let us assume that the photospheric emission has a blackbody spectrum at temperature $T_{\rm s}$, meaning that
\begin{equation}
    I_{\mr{s},\nu} = D_{\rm s} B_\nu(T_{\rm s}),\ \ \pi I_{\rm s} = D_{\rm s} \sigma_{\rm SB}T_{\rm s}^4,
\end{equation}
where $B_\nu(T)=2h\nu^3c^{-2}(\mr{e}^{h\nu/\kB T} - 1)^{-1}$ ($h$ being the Planck constant) is the Planck function and we have defined a dilution factor $D_{\rm s}$ based on eq. (\ref{eq:Ls}),
\begin{equation}
    D_{\rm s} \equiv {L_{\rm s}\over 4\pi R_{\rm s}^2 \sigma_{\rm SB} T_{\rm s}^4} = 4\lrb{3\kapa\over \kaps}^{3/2} < 1.
\end{equation}

\begin{figure}
  \centering
\includegraphics[width = 0.49\textwidth]{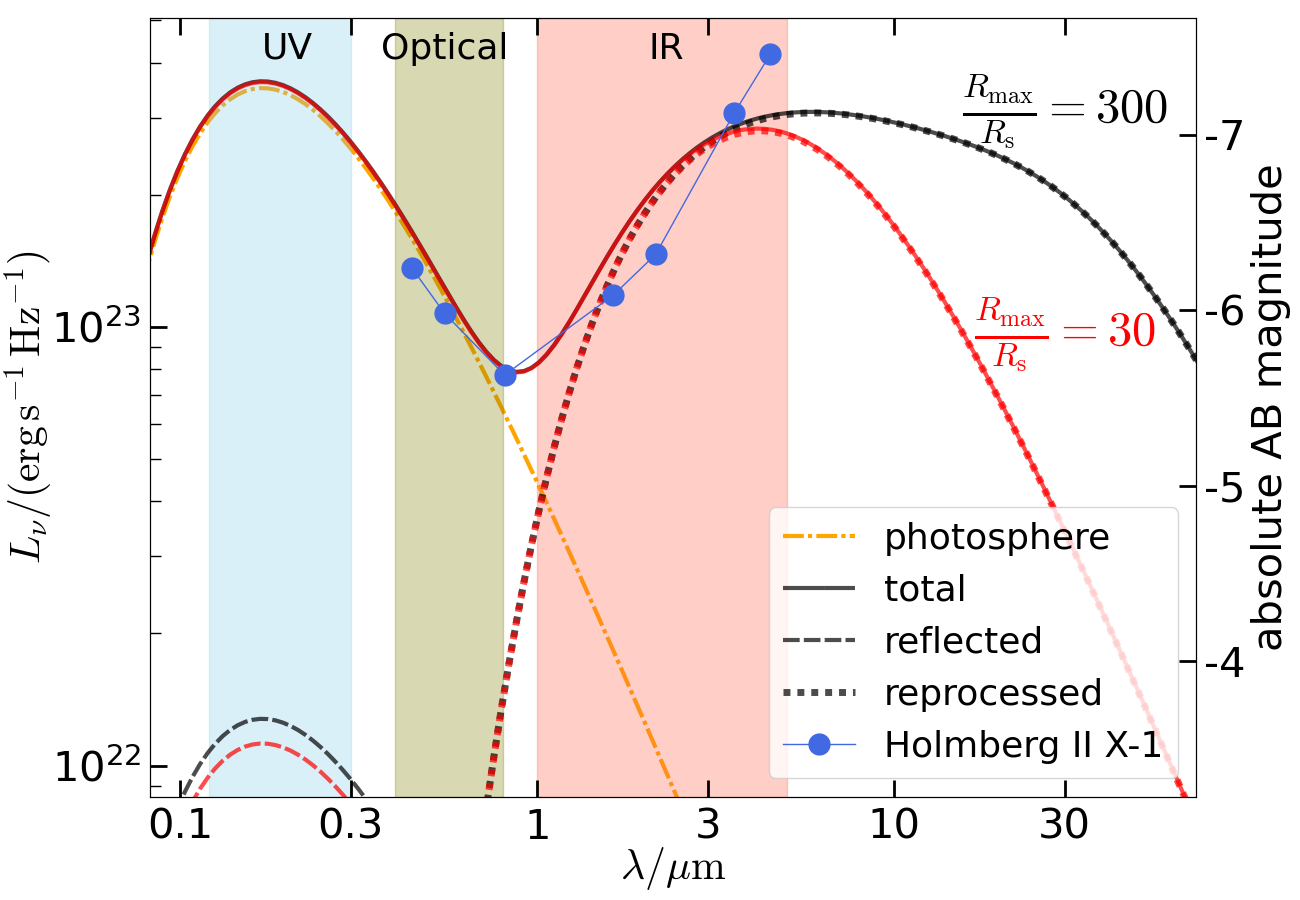}
\caption{Spectra from a spherical photosphere and an optically thick, geometrically thin circum-binary sheet in the equatorial plane, for two cases of different outer radii $R_{\rm max}/R_{\rm s} = 300$ (black) and 30 (red). For both cases, the inner radius of the equatorial sheet is $R_{\rm min}/R_{\rm s}=4$ and the scattering photospheric radius is $R_{\rm s}=0.5\rm\, AU$. The emission emerging from the photosphere has luminosity $L_{\rm s}=10^{39}\rm\, erg\,s^{-1}$ and is assumed to be a blackbody at temperature $T_{\rm s}=3\times 10^4\rm\, K$ (orange dash-dotted line). The equatorial sheet has two emission components --- reflected component from the photo-ionized hot surface (dashed lines) and reprocessed component from the deep interior (dotted lines) --- and their relative contribution is given by the albedo $\varpi=0.4$. The observer's viewing angle wrt. to the rotation axis is $\theobs=\pi/6$. For comparison, we show the spectral energy distribution of Holmberg II X-1 (a bright ULX, see text for references) where the IR excess is consistent with reprocessing by circum-binary material.
}\label{fig:sheet_reflection}
\end{figure}

At radii $R\ll R_{\rm max}$, only a thin surface layer with $\Delta H/H\ll 1$ is photo-ionized and the gas density in the surface layer is much less than that at the equatorial plane. Depending on the detailed vertical density profile (which determines the ratio between absorption and scattering opacity at different heights), this surface layer has an albedo $0<\varpi<1$, where the upper (or lower) limit corresponds to a scattering (or absorption) dominated surface layer. The net result is that a fraction $1-\varpi$ of the flux is absorbed by the CBO, and we assume this amount of energy is re-emitted as a blackbody at the local effective temperature given by
\begin{equation}
    \sigma_{\rm SB} T_{\rm eff}^4(R) =  (1-\varpi)F(R),
\end{equation}
whereas the rest of the incident flux $\varpi F(R)$ is reflected away from the CBO. Physically, the albedo $\varpi$ is expected to be a (mildly) decreasing function of radius, because the fractional height $\Delta H/H$ of the photo-ionized layer increases with radius and this leads to a decrease in the ionization parameter (which is defined as the ratio between the incident photon number density and the gas number density). For the purpose of obtaining rough estimates for the reprocessed emission, we simply adopt a constant $\varpi$ throughout the entire range of radii.

In the limit $R\gg R_{\rm s}$, we have $F\propto R^{-3}$ and the effective temperature of the re-emission scales as $T_{\rm eff}\propto R^{-3/4}$. Since $T_{\rm eff}$ is proportional to the peak emission frequency from each annulus, this generates a power-law spectrum $L_\nu \propto \d L/\d T_{\rm eff}\propto \d (R^2F)/\d T_{\rm eff}\propto \nu^{1/3}$, which is much shallower than the Rayleigh-Jeans slope. This power-law truncates at radius $R_{\rm max}$ (the shape of the spectrum at $\nu\ll \nu_{\rm min}$ depends on the emission from the photo-ionized gas at $R\gtrsim R_{\rm max}$), which means the spectrum below $\nu_{\rm min} \simeq (R_{\rm max}/R_{\rm s})^{-3/4} k_{\rm B}T_{\rm s}/h$ will be Rayleigh-Jeans-like (however, free-free emission from the photo-ionized gas will dominate at $\nu\lll \nu_{\rm min}$, which we do not consider here). Note that, if the equatorial outflow has a flared structure (i.e. height-to-radius ratio increasing with radius) or the albedo $\varpi$ decreases with radius, then the re-emission at longer wavelengths is enhanced compared to our calculation \citep{chiang97_passive_disk}.

For a line-of-sight inclination angle of $\theobs$ wrt. the rotational axis of the system, the isotropic equivalent specific luminosity at frequency $\nu$ is given by
\begin{equation}\label{eq:reflection_spectrum}
\begin{split}
    L_\nu =&\, 4\pi^2 R_{\rm s}^2 I_{\mr{s},\nu} +\\
    &\,4\pi \cos\theobs\int_{R_{\rm min}}^{R_{\rm eq}} \left[B_\nu(T_{\rm eff}) + {\varpi F(R)\over \sigma_{\rm SB}T_{\rm s}^4} B_\nu(T_{\rm s})\right] 2\pi R\d R,
\end{split}
\end{equation}
where the first term is the contribution from the wind photosphere itself, and the second term has a cold thermal ($T_{\rm eff}$) component from deeper layers close to the mid-plane and a hot reflected component $(T_{\rm s})$ from the photo-ionized surface layer. Then equations (\ref{eq:reflection_flux}) and (\ref{eq:reflection_spectrum}) can be re-written in the following dimensionless forms
\begin{equation}
    \tilde{F}\left(x\right) \equiv {F(R)\over \pi I_{\rm s}} = {2\over \pi} \int_{x^{-1}}^1 \d \tilde{\mu} {(x\tilde{\mu} - 1)\sqrt{1-\tilde{\mu}^2} \over (x^2 + 1 - 2x\tilde{\mu})^2},
\end{equation}
and
\begin{equation}
\begin{split}
    \tilde{L}_{\nu} \equiv &\, {L_\nu \over 4\pi^2 R_{\rm s}^2 I_{\mr{s},\nu}} = 1+ 2\cos\theobs \times \\
   &\, \int_{x_{\rm min}}^{x_{\rm max}} \d x\, x  \left[D_{\rm s}^{-1}{\mr{e}^{h\nu/\kB T_{\rm s}} - 1 \over \mr{e}^{h\nu/\kB T_{\rm eff}} - 1} + \varpi D_{\rm s} \tilde F(x) \right], 
\end{split}
\end{equation}
where $x = R/R_{\rm s}$, $x_{\rm min}=R_{\rm min}/R_{\rm s}$, $x_{\rm max} =R_{\rm max}/R_{\rm s}$, and $T_{\rm eff}(x)/T_{\rm s} = [(1-\varpi) D_{\rm s}\tilde F(x)]^{1/4}$.

The spectra of the reprocessing system are shown in Fig. \ref{fig:sheet_reflection} for two cases of different outer radii $R_{\rm max}/R_{\rm s} = 300$ and $30$. We adopt the following parameters: wind luminosity $L_{\rm s}=10^{39}\rm\, erg\,s^{-1}$, source temperature $T_{\rm s}=3\times 10^4\rm\, K$, photospheric radius $R_{\rm s} = 0.5\rm\, AU$, inner CBO radius $R_{\rm min}=4R_{\rm s}$, albedo of the photo-ionized layer $\varpi=0.4$, and observer's viewing angle $\theobs = \pi/6$. These parameters are motivated by the consideration of a binary system with mass-transfer rate of the order $10^{-3}\,\msunyr$. We find that the IR luminosity from such a system is $L_{\rm IR}\sim 5\times 10^{37}\rm\, erg\,s^{-1}$. For lower $|\dot{M}_1|$, the source temperature $T_{\rm s}$ will be higher and the photospheric radius smaller, so the reprocessed IR luminosity will be lower. We note that the IR luminosity due to reprocessing is likely brighter than that generated by internal shocks within the CBO \citep{pejcha16_L2_stream_shape}
\begin{equation}
\begin{split}
    L_{\rm sh} & \sim 0.03 \fL2  {GM |\dot{M}_1|\over a}\\
    &\simeq 3\times10^{35}\mr{\,erg\,s^{-1}} {\fL2 |\dot{M}_1|\over 10^{-3}\,\msunyr} {M\over 10\Msun} {100\Rsun\over a}.
\end{split}
\end{equation}

Our model prediction is compared to the SED of a luminous ULX (with apparent X-ray luminosity $L_{\rm X}\sim 10^{40}\rm\, erg\,s^{-1}$) in the nearby dwarf star-forming galaxy Holmberg II, Holmberg II X-1 \citep[e.g.,][]{kaaret04_holmbergIIX1_He_line, goad06_holmbergIIX1_xray_timing, berghea10_holmbergIIX1_MIR, tao12_holmbergIIX1_optical, walton15_holmbergIIX1_X-ray}. This ULX has bright optical ($U,B,I$ bands), near-IR (\textit{Keck} $H$ and $K_{\rm s}$ bands) and mid-IR (warm \textit{Spitzer} IRAC 3.5 and 4.5$\,\mu \mr{m}$ bands) counterparts \citep{tao11_opt_counterparts, heida14_NIR_counterparts, lau17_MIR_counterparts, lau19_ULX_MIR}. The median is used if the flux at a given band has significant time variability. The IR flux greatly exceeds the extrapolation from the optical bands and has been interpreted as the emission from circum-stellar dust or a supergiant Be star by \citet{lau17_MIR_counterparts, lau19_ULX_MIR}. Here, our model provides a physical explanation of the bright IR emission being due to reprocessing by a CBO from L2 mass loss. We do not provide a statistical fit to the SED of Holmberg II X-1, because the photometry in the near-IR (or mid-IR) was carried out with instruments with angular resolution of about $1''$ (or a few $''$), and the contribution from diffuse emission and noise in the adopted aperture (of physical radii of 10 to 30 pc at the source distance) may be significant. 


We also note that the brightest Galactic micro-quasar SS433 \citep[see][for a review]{fabrika04_SS433} is undergoing mass transfer at a rate that is estimated to be of the order $10^{-4}\, \msunyr$ based on the mass outflow rate inferred from the infrared spectrum \citep{shkovskii81_SS433_Mdot, vandenheuvel81_SS433_Mdot} and its binary separation is $a\simeq 60\,\Rsun [(M_1+M_2)/20\Msun]^{1/2}$ for an orbital period of 13.1 d. There is strong evidence for the existence of CBO in this system\footnote{The evidence includes (1) that the stationary H$\alpha$ lines (and other recombination lines) have two narrow components that are blue-shifted and red-shifted with half-separation of about $200\rm\, km\,s^{-1}$ independent of the orbital revolution whereas the inner regions of the accretion disk undergoes eclipses in each orbit --- these two narrow components must come from circum-binary gas \citep{blundell08_ss433_eq_outflow}; 
(2) the spatially resolved images of radio (free-free) emission shows outflowing gas in the equatorial plane roughly perpendicular to the jets \citep{paragi99_SS433_CBO_radio_image, blundell01_ss433_eq_outflow};
(3) near-IR interferometric observations shows that the low-velocity core of the stationary Br$\gamma$ line is emitted by gas moving perpendicular to the jet directions \citep{waisberg19_ss433_eq_outflow}.} \citep[as initially suggested by][]{filippenko88_ss433_disk_emission}.
However, since SS433 is close to the Galactic plane with high (and uncertain) dust extinction $A_{\rm V}\sim 8\rm\, mag$ \citep{wagner86_SS433_Av}, we do not attempt to model the SED of this source here.

Our model can also be applied to systems shortly before one of the stars undergoes a SN explosion. If a pre-SN system has strong L2 mass loss, 
reprocessing of the UV and soft X-ray emission from the disk wind by the equatorial outflow should generate bright near-IR excess at the level of $\sim 10\%$ of the accretor's Eddington luminosity. This provides a possible explanation of the photometric detections of ``cool'' progenitor systems of some SNe \citep{smartt15_preSN_imaging}, because the equatorial outflow has a much larger surface area than that of the pre-explosion star. For instance, the Type Ib SN 2019yvr had archival multi-band pre-explosion images at its position showing bright emission with $\nu L_\nu \simeq 7\times 10^{37}\rm\, erg\,s^{-1}$ at wavelength $\lambda= 0.8\, \mu\rm m$ \citep{kilpatrick21_2019yvr_preSN_image, sun21_SN2019yvr}. This is consistent with reprocessing by a CBO but difficult to explain by a single, hydrogen-poor star model. Moreover, SN 2019yvr transitioned into a Type IIn at late time with narrow H$\alpha$ emission line, X-ray and radio emission from shock interaction with circum-stellar material, further supporting our picture.

\section{Discussion}\label{sec:discussion}

In this section, we discuss the limitations of our model, which may be improved in the future.

(1) The boundary between $\fL2\ll 1$ and $\fL2\sim 1$ in the $|\dot{M}_1|$-$a$ plane (Fig. \ref{fig:fL2outer}) is affected by many factors that are explicitly contained in our model description. In the Appendix, we show how the results depend on the gas composition, which affects the Rosseland-mean opacity and hence the radiative cooling rate $Q_{\rm rad}$. Generally, at lower opacities (e.g., for a hydrogen poor composition), higher mass-transfer rates are needed to trigger L2 mass loss. The results also depend on the dimensionless viscosity parameter $\alpha$ (fixed at fiducial value of $0.1$) and mass ratio $q=M_2/M_1$ (fixed at fiducial value of 0.5). We do not explore the entire multi-dimensional parameter space here in this paper. Instead, we make our source code public\footnote{\href{https://github.com/wenbinlu/L2massloss.git}{https://github.com/wenbinlu/L2massloss.git}}, so it can be used to calculate the L2 mass loss fraction for any set of parameters.

(2) The outer disk near radius $\Rd$ is described by a one-zone model based on global mass/energy conservation. Such a one-zone model is only a crude approximation of the full 3D structure of the accretion flow, and the quantitative results can only be trusted to within a factor of a few. More realistically, because the gas at different radii experiences different heating and cooling rates, the Bernoulli number is a function of radius and this radial dependence determines which part of the disk is susceptible to wind or L2 mass loss. Moreover, if we consider the gas profile in the vertical direction, at radii where radiative cooling is inefficient, the gas at higher latitudes tends to have a higher Bernoulli number and hence is escaping the disk more easily \citep{stone99_ADAF}. Furthermore, a fraction of the gas is located near the tidal truncation radius (which is slightly smaller than the volume-equivalent Roche-lobe radius of the secondary, $R_{\rm v,2}$) and their orbits are significantly non-circular. The assumption of Keplerian rotation breaks down. These multi-dimensional effects are better captured in (more expensive) numerical simulations \citep[e.g.,][]{macleod18_L2_loss_binary_merger}, which will be explored in the future.

(3) Our radiative cooling term $Q_{\rm rad}$ (eq. \ref{eq:Qrad}) considers photon diffusion in a uniform gas, whereas more realistically, the gas density profile is likely highly inhomogeneous such that photons try to escape through the lower density regions along the paths of least resistance \citep{blaes11_vertical_transport, jiang14_rad_MHD}. If radiative cooling is more efficient than in our model, then the gas Bernoulli number is reduced and hence a higher mass-transfer rate is needed to trigger L2 mass loss.


(4) We focus on the case of stable mass transfer. The donor loses its envelope on a timescale much longer than the dynamical time of $\Omega^{-1}$ ($\Omega$ being the binary orbital frequency). In the case of dynamically unstable mass transfer (e.g., right before a merger event), the assumption of corotation breaks down and our model provides a poor description for the hydrodynamics of the system. At extremely high mass-transfer rates $|\dot{M}_1| \gtrsim 10^{-2}\, \msunyr$, a fraction of the shock-heated gas with the highest Bernoulli number is likely lost on a dynamical time. This fraction of gas is unable to form an accretion disk and hence does not contribute to viscous heating. It is also likely that some of the high-Bernoulli-number gas is lost through the L3 nozzle. We speculate that the mass loss from the system is still concentrated near the equatorial plane, but the height-to-radius ratio of the CBO likely increases with the mass-transfer rate.

(5) We have restricted our analysis to the case where the accretor's radius is less than the circularization radius of the incoming stream from the L1 nozzle ($R_2 < \Rd$). In many binary systems (e.g., double main-sequence or double white-dwarf), the two stars have similar radii such that $R_2 > \Rd$, and hence the stream directly hits the surface of the accretor. At low mass-transfer rates, the mass transfer is conservative such that $\dot{M}_2 = -\dot{M}_1$ (up to the point where the accretor is spun to near the critical rotation rate). However, for high mass-transfer rates such that $GM_2\dot{M}_1/R_2 > L_{\rm Edd,2}$, the shocked gas is unable to radiatively cool ($Q_{\rm rad}^-\ll Q_{\rm sh}^+$). The strong radiation pressure potentially pushes a fraction of the incoming gas (as well as some material originally from the accretor) away from the system through the L2 nozzle, and the rest of the gas settles down on the surface of the accretor. Thus, we still expect the formation of an equatorially concentrated CBO. The consequence of the rapid angular momentum loss associated with the L2 outflow is that the system is more likely to undergo a violent merger than in the case of conservative mass transfer. The merger ejecta interacts with the CBO and generates a bright transient in the optical band \citep{metzger17_csm_interaction}.


(6) We do not specify the physical origin of the high mass-transfer rates considered in this work. Due to their short lifetime ($10^3$ to $10^4\rm\, yr$) and the requirement of massive binary initial conditions, these systems are very rare in the Universe, with cosmic number density of the order $10^8\rm\, Gpc^{-3}$ or less. Despite their rarity, such systems may produce some of the most interesting sources in high-energy astrophysics: compact object mergers, stellar mergers, micro-quasars/ULXes, and interaction-powered SNe. Future binary population modeling is needed to identify possible evolutionary pathways that lead to the high mass-transfer rates.

\section{Summary}\label{sec:summary}
This paper considers the hydrodynamics of semi-detached binary systems where the donor transfers mass to the companion via Roche-lobe overflow.
We construct a physical model for the accretion disk around the companion. It is proposed that, at sufficiently high mass transfer rates such that the accretion flow becomes geometrically thick (or advection-dominated) near the outer disk radius, a large fraction of the transferred mass is lost through the L2 point. This is based on the physical intuition that losing mass from the L2 point is energetically favorable over lifting material to infinity by a fast wind.

Our model predicts the fraction of the transferred mass that is directly lost through the L2 nozzle from the outer disk $f_{\rm L2}^{\rm outer}$, which becomes of the order unity when $|\dot M_1|\gtrsim \mbox{few}\times 10^{-4}\,\msunyr$. Our model is tentatively supported by the (although inviscid) hydrodynamic numerical simulations by \citet{macleod18_L2_loss_common_envelope, macleod18_L2_loss_binary_merger}, who studied binary systems undergoing unstable mass transfer (at a rate of the order $10^{-2}\,\msunyr$ or higher) and identified mass loss through the L2 point (as well as the L3 point). Future MHD simulations taking into account viscous accretion are needed to further test our model.

At lower mass-transfer rates $|\dot M_1|\lesssim 10^{-4}\,\msunyr$, when direct L2 mass loss from the outer disk is negligible ($\fL2^{\rm outer}\approx 0$), the inner disk can launch a strong outflow near the spherization radius $R_{\rm sph}$ where the accretion luminosity approaches the Eddington limit of the accretor, provided that the size of the accretor is small (e.g., in the case of a compact object). In this case, the majority of the disk wind directly leaves the system at a typical speed $v_{\rm w}\sim \sqrt{GM_2/\Rsph}$, whereas a small fraction of (or the slowest part of) the wind can still be captured by the equipotential surfaces passing near the L2 point. We estimate the L2 mass loss fraction from the inner disk to be $\fL2^{\rm inner}\sim |\Phi_{\rm L2}|/(GM_2/\Rsph)$ and we find it to be small: $\fL2^{\rm inner}$ is at most a few percent at low mass-transfer rates $|\dot M_1|\lesssim 10^{-4}\,\msunyr$.


Material leaving the binary system via the L2 point forms a circum-binary outflow (CBO), although it is also possible to produce a decretion disk for a narrow range of binary mass ratio $0.8\lesssim q\lesssim 1.3$ \citep{shu79_L2_stream}. Due to the large lever-arm of the L2 point, a large L2 mass loss fraction may cause rapid angular momentum loss, which tends to shrink the orbital separation, leading to shorter orbital periods or potentially unstable mass transfer.
The effects on binary evolution and merger rates of gravitational wave sources will be explored in a companion paper with $\mathtt{MESA}$ \citep{paxton19_mesa} simulations.

A key signature of a system undergoing L2 mass loss is that the UV/optical emission from the hot accretion disk wind is reprocessed by the CBO into the IR bands. This produces an IR luminosity that can reach about $10\%$ of the Eddington luminosity of the accretor. Thus, we encourage taking high-resolution images of nearby ULXes with the James Webb Space Telescope to identify possible IR counterparts and measure their IR fluxes. Another implication of our model is that, if a binary had L2 mass loss before the SN explosion of one of the stars, then the pre-SN system (if detected in archival images) should have an IR excess due to reprocessing by the CBO. This provides a possible explanation of the ``cool progenitor'' of Type Ib SN 2019yvr \citep{kilpatrick21_2019yvr_preSN_image}. 


\section*{Data Availability}
The data underlying this article will be shared on reasonable request to the corresponding author.

\section*{Acknowledgement}
We thank Nadia Zakamska, Jim Stone, Alexey Bobrick and Pablo Marchant for useful conversations. We are grateful for the careful read and detailed comments made by the referee, Christopher Tout. WL was supported by the David and Ellen Lee Fellowship at California Institute of Technology and the Lyman Spitzer, Jr. Fellowship at Princeton University. This project has received funding from the European Union's Horizon 2020 research and innovation programme under the Marie Sklodowska-Curie grant agreement No 836751.

{\small
\bibliographystyle{mnras}
\bibliography{refs}
}

\appendix
\section{Effects of Opacity}
\begin{figure}
  \centering
\includegraphics[width = 0.49\textwidth]{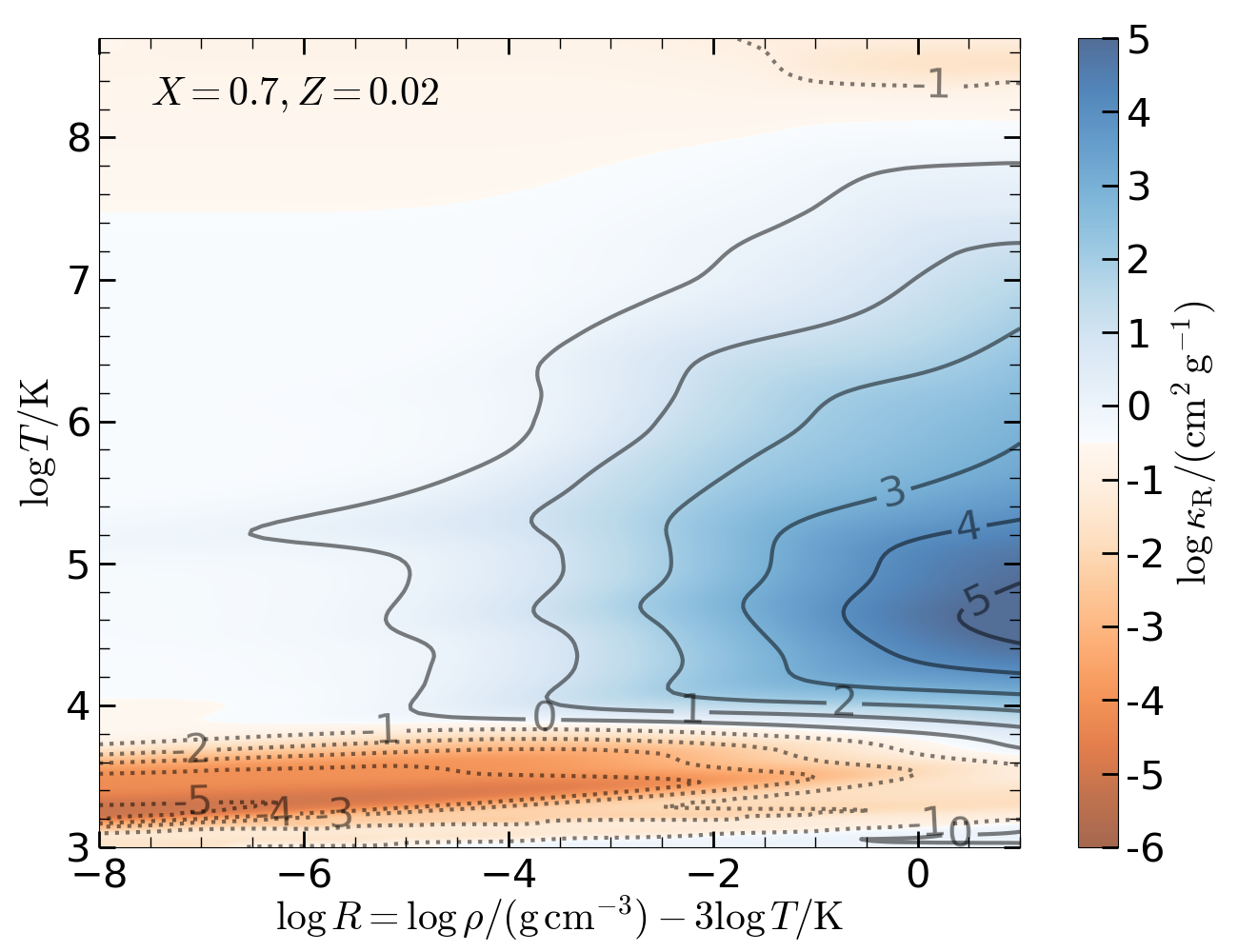}
\includegraphics[width = 0.49\textwidth]{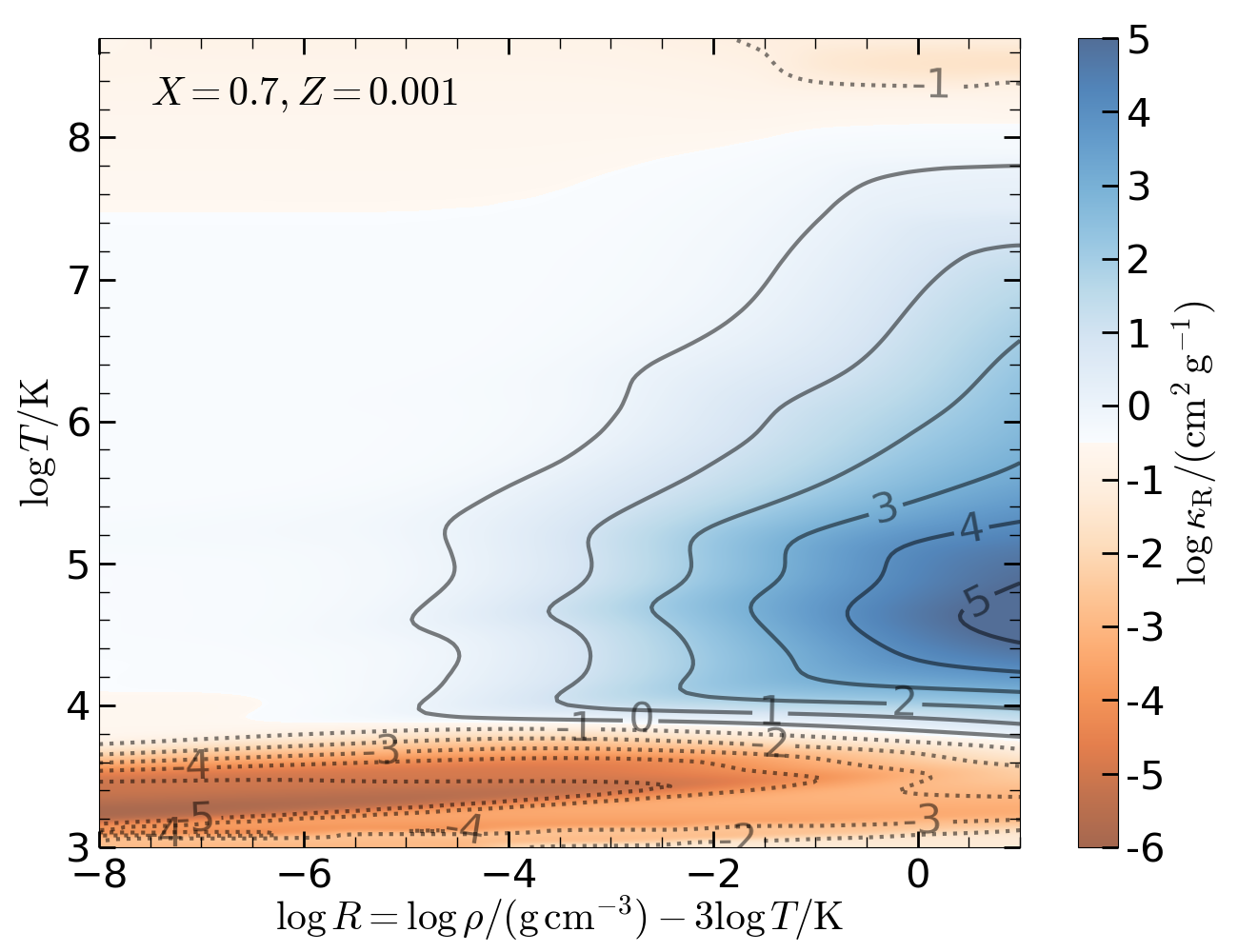}
\includegraphics[width = 0.49\textwidth]{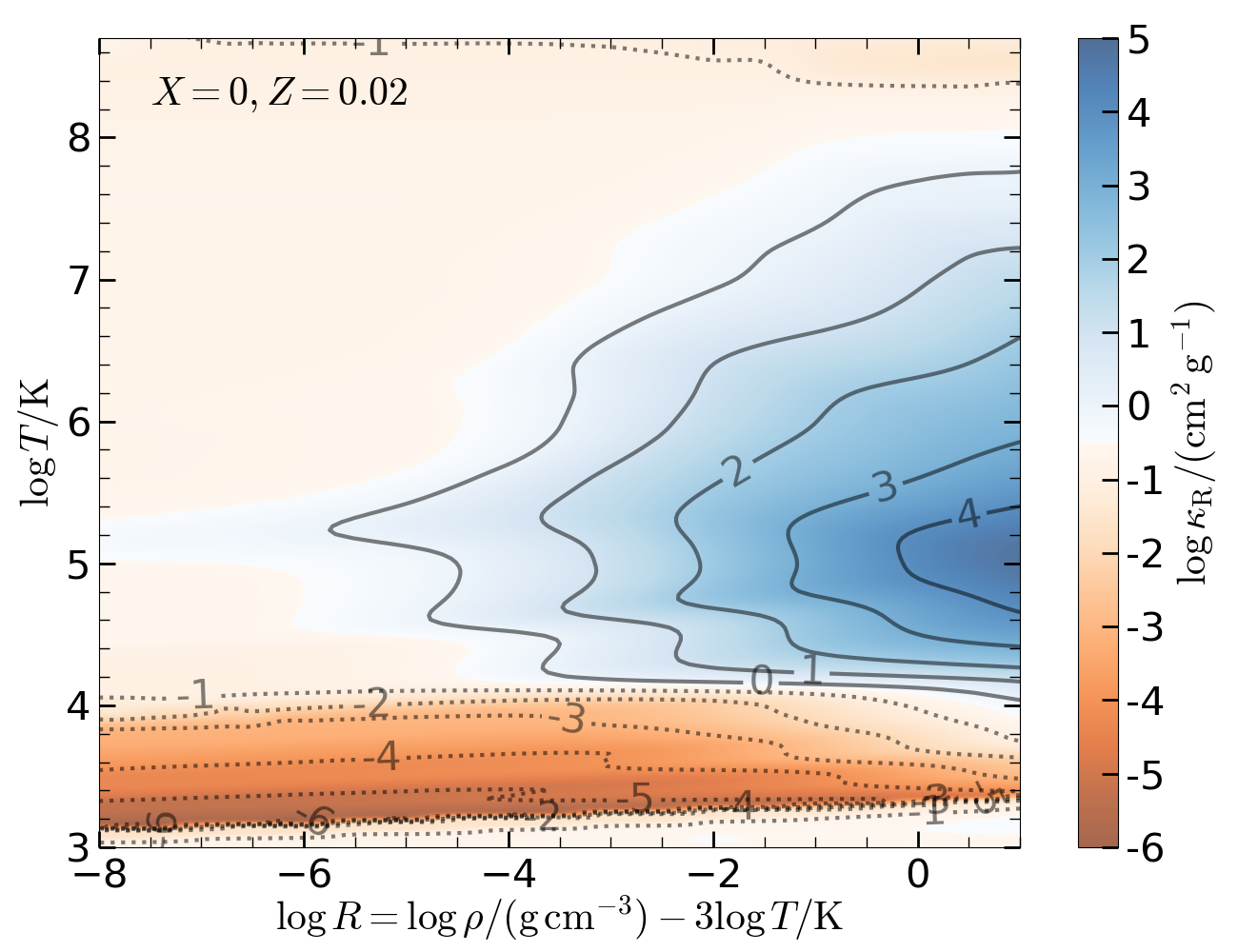}
\caption{Rosseland-mean opacity tables corresponding to three different gas compositions: H-rich solar metallicity ($X=0.7, Z=0.02$, upper), H-rich low metallicity ($X=0.7, Z=0.001$, middle), and H-poor solar metallicity ($X=0, Z=0.02$, bottom panel). Compared to the standard ``H-rich $Z_\odot$'' case, the low-metallicity case lacks the iron opacity bump near $T\sim 2\times 10^5\rm\, K$, and the H-poor case has lower electron scattering opacity and lacks the hydrogen opacity bump near $T\sim 1\times 10^4\rm\, K$. 
}\label{fig:kappa}
\end{figure}

In this Appendix, we show the dependence of our results on the Rosseland-mean opacity, which affects the cooling of the disk gas by radiative diffusion in the vertical direction. We consider a H-poor solar-metallicity $(X=0, Z=0.02)$ gas composition, as well as a H-rich low-metallicity ($X=0.7, Z=0.001$) case. The main differences from our fiducial case $(X=0.7, Z=0.02)$ are that the H-poor case has a lower scattering opacity and the low-metallicity case has a weaker iron opacity bump, and the consequence is that L2 mass loss occurs at a higher mass-transfer rate (for a fixed binary separation and component masses). The opacity tables $\kappa(\rho, T)$ for all three cases are shown in Fig. \ref{fig:kappa}. We use the boundary values at the same temperature when $(\rho, T)$ is beyond the boundaries of the given table. This only occurs for wide binary separation $a\gtrsim10^3\Rsun$ and low mass-transfer rate $|\dot{M}_1|\lesssim 10^{-5}\,\msunyr$, and our main results are unaffected.

\begin{figure*}
  \centering
\includegraphics[width = 0.49\textwidth]{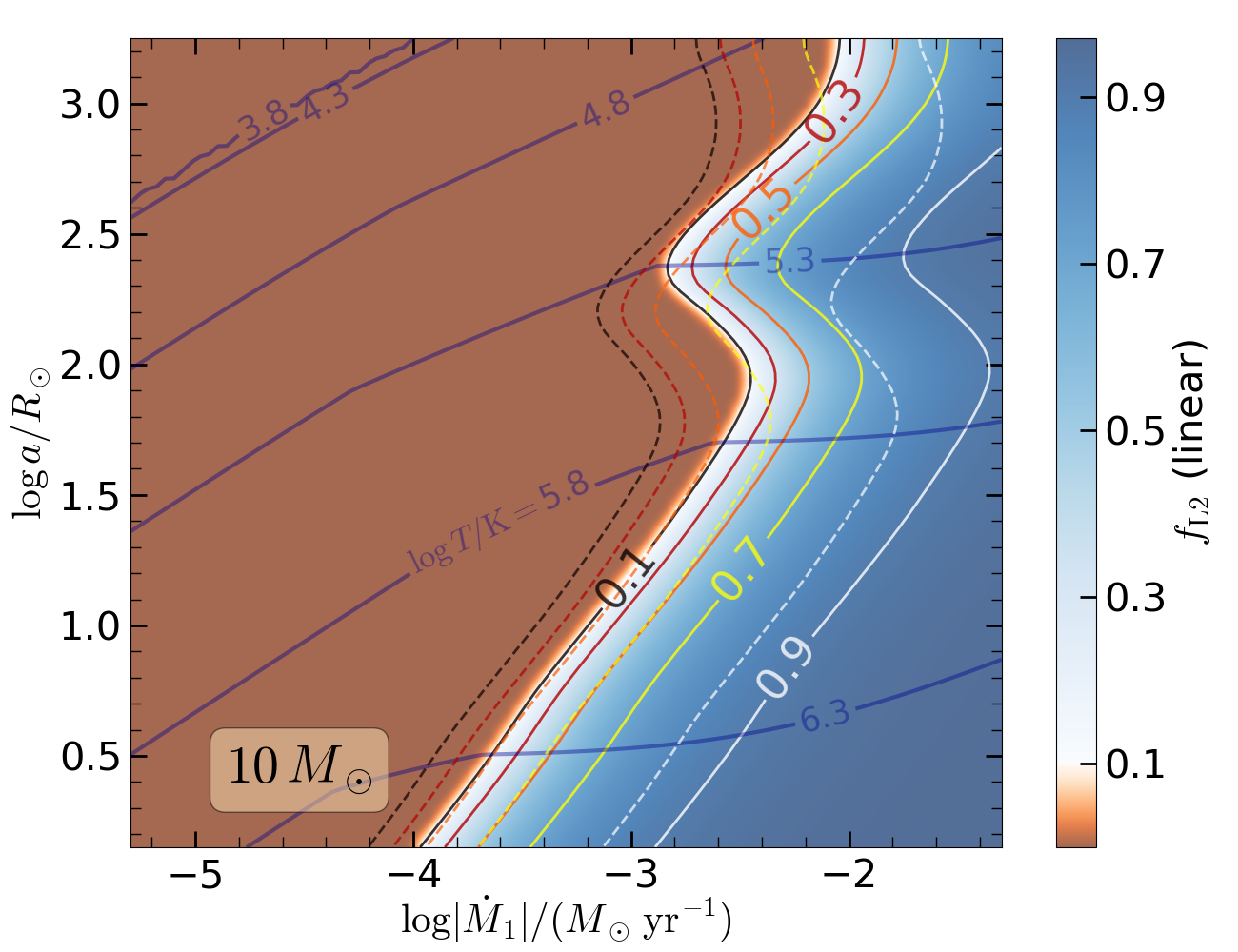}
\includegraphics[width = 0.49\textwidth]{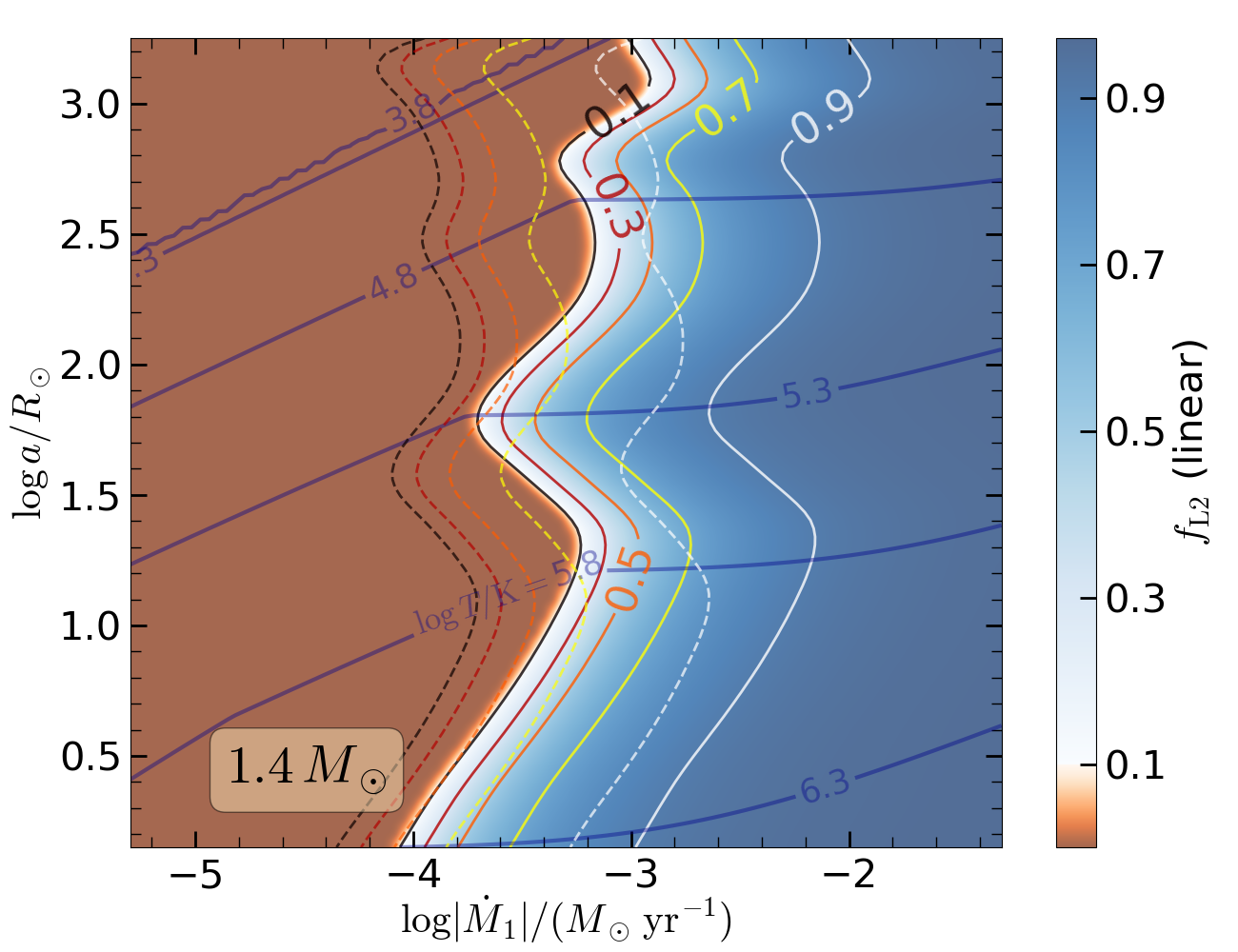}
\caption{The color-shading and solid contour lines show the L2 mass loss fraction from the outer disk $\fL2^{\rm outer}$ (in linear scale) for H-poor solar-metallicity gas composition ($X=0, Z=0.02$). The left panel is for secondary mass $M_2 = 10\Msun$ (massive star or BH), and the right panel is for $M_2 = 1.4\Msun$ (NS). For both panels, we have fixed the mass ratio $q = M_2/M_1 = 0.5$ and viscosity parameter $\alpha=0.1$, and we use $\mu_{\rm g}=4/3$ (mean molecular weight for fully ionized helium, see text below eq. \ref{eq:disk_height}). Compared to the fiducial case of H-rich solar-metallicity gas (shown by the dashed contours), we find that the main effect of lower hydrogen mass fraction is to reduce the scattering opacity as well as the hydrogen opacity bump near $T\sim 1\times 10^4\rm\, K$; this increases the Eddington luminosity of the accretor, so higher mass-transfer rates are needed to trigger L2 mass loss. The effect of the reduced hydrogen opacity bump is only seen on the right panel ($M_2 = 1.4\Msun$) for wide binary separations $a\sim 10^3\Rsun$.
}\label{fig:fL2out_Hpoor}
\end{figure*}

\begin{figure*}
  \centering
\includegraphics[width = 0.49\textwidth]{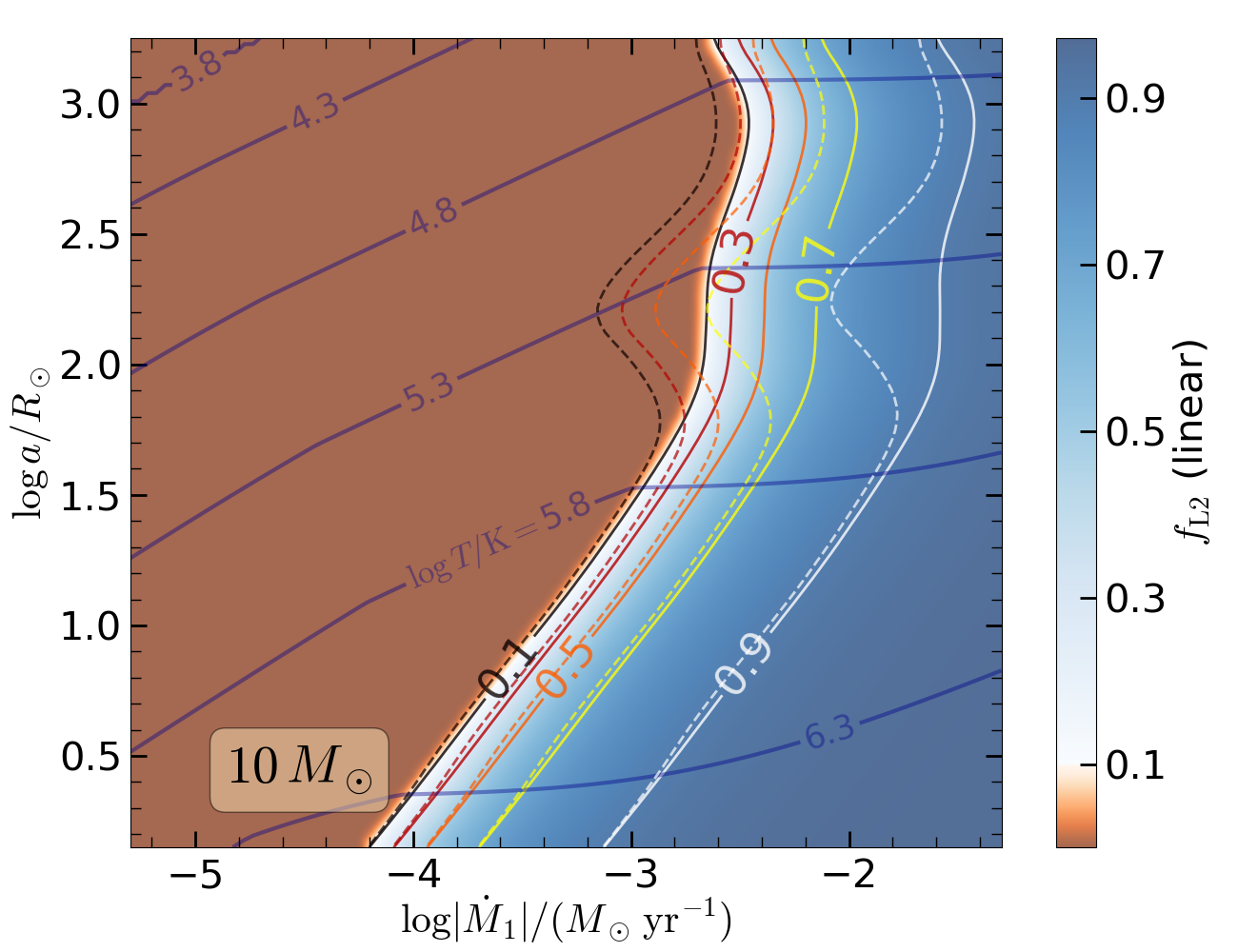}
\includegraphics[width = 0.49\textwidth]{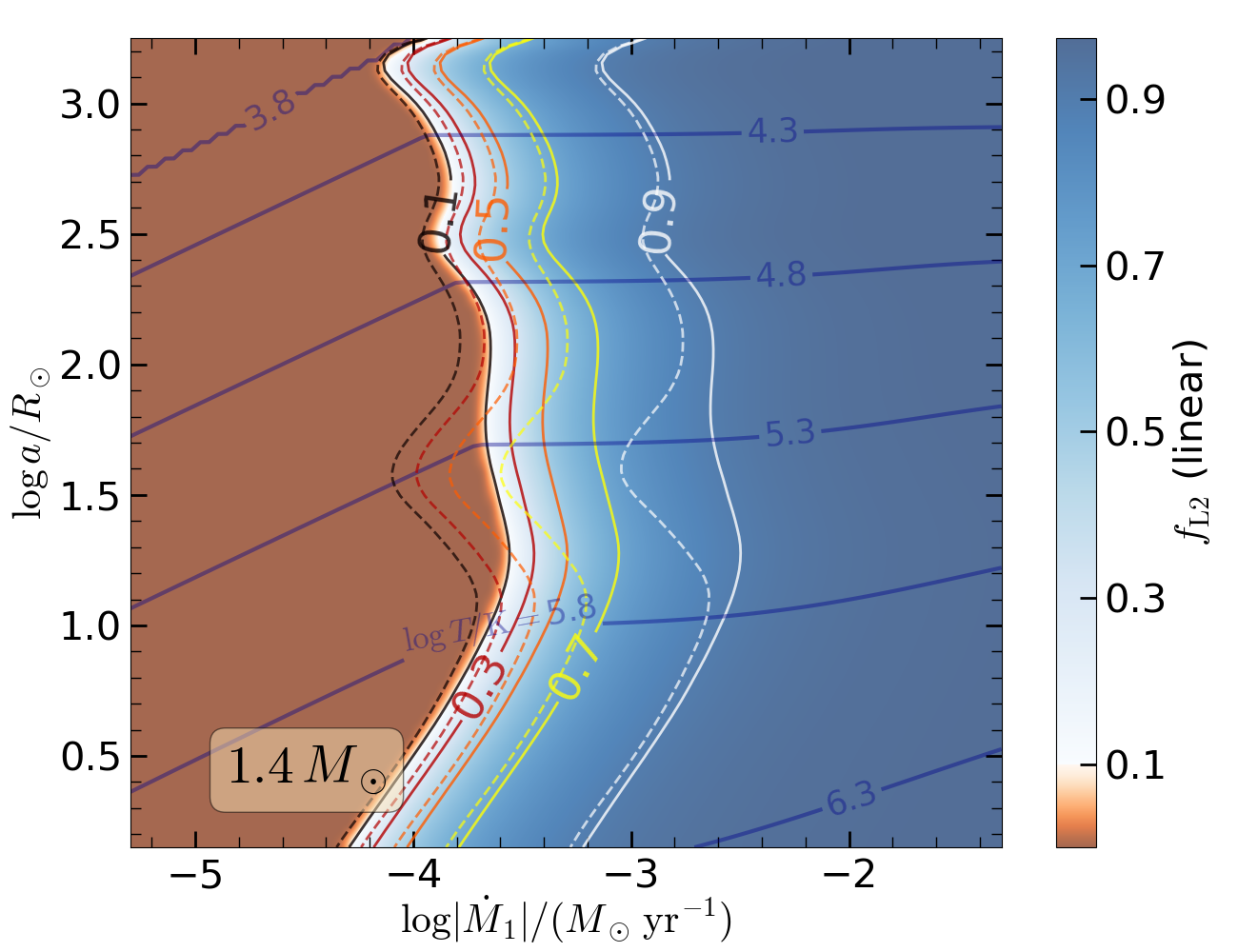}
\caption{The color-shading and solid contour lines show the L2 mass loss fraction from the outer disk $\fL2^{\rm outer}$ (in linear scale) for H-rich low-metallicity gas composition ($X=0.7, Z=0.001$). The left panel is for secondary mass $M_2 = 10\Msun$ (massive star or BH), and the right panel is for $M_2 = 1.4\Msun$ (NS). For both panels, we have fixed the mass ratio $q = M_2/M_1 = 0.5$ and viscosity parameter $\alpha=0.1$. Compared to the fiducial case of H-rich solar-metallicity gas (shown by the dashed contours), we find that the main effect of lower metallicity is to reduce the iron opacity bump near $T\sim 2\times 10^5\rm\, K$; this increases the Eddington luminosity of the accretor, so higher mass-transfer rates are needed to trigger L2 mass loss.
}\label{fig:fL2out_Hrich_lowZ}
\end{figure*}

Then, we show the results of the L2 mass loss fraction from the outer disk $\fL2^{\rm outer}(|\dot{M}_1|, a)$ as a function of mass-transfer rate and binary separation in Fig. \ref{fig:fL2out_Hpoor} (for H-poor solar metallicity) and in Fig. \ref{fig:fL2out_Hrich_lowZ} (for H-rich low metallicity). In each of the figures, we compare the result with that from in our fiducial case of H-rich solar metallicity composition.

\end{document}